\documentstyle[preprint,aps,eqsecnum,epsfig]{revtex}

\begin{document}
\preprint{CMU-HEP-97-05; DOR-ER/40682-130; LPTHE-97-05; PITT-97-10; UCSBTH-97-03}
\draft
\title{\bf Scalar Field Dynamics in Friedman Robertson Walker Spacetimes}
\author{{\bf D. Boyanovsky$^{(a)}$, D. Cormier$^{(b)}$, H. J. de Vega$^{(c)}$,
R. Holman$^{(b)}$, A. Singh$^{(d)}$ and M. Srednicki$^{(d)}$}}
\address
{{\it (a)  Department of Physics and Astronomy, University of
Pittsburgh, Pittsburgh, PA. 15260, U.S.A.} \\
{\it (b) Department of Physics, Carnegie Mellon University, Pittsburgh,
PA. 15213, U. S. A.} \\
{\it (c)  Laboratoire de Physique Th\'eorique et Hautes Energies$^{[*]}$
Universit\'e Pierre et Marie Curie (Paris VI), 
Tour 16, 1er. \'etage, 4, Place Jussieu 75252 Paris, Cedex 05, France} \\
{\it (d) Department of Physics, University of
California, Santa Barbara, CA 93106, U. S. A.}}
\date{February 1997}
\maketitle
\begin{abstract}
We study the non-linear dynamics of quantum fields in matter and radiation
dominated universes, using the non-equilibrium field theory approach combined
with the non-perturbative Hartree and the large $N$ approximations.  We examine
the phenomenon of explosive particle production due to spinodal instabilities
and parametric amplification in expanding universes with and without symmetry
breaking.  For a variety of initial conditions, we compute the evolution of the
inflaton, its quantum fluctuations, and the equation of state.  We find
explosive growth of quantum fluctuations, although particle production
is somewhat sensitive to the expansion of the universe.  
In the large $N$ limit for symmetry breaking
scenarios, we determine generic late time solutions for any flat
Friedman-Robertson-Walker cosmology.  We also present a complete
and numerically implementable renormalization scheme for the
equation of motion and the energy momentum tensor in flat FRW
cosmologies. In this scheme the renormalization constants are 
 independent  of time and of the initial conditions. 
\end{abstract}

\section{Introduction}

Over the last fifteen years, there has been a sustained effort to study the
evolution of scalar fields within the context of the early universe.  These
efforts have been largely fueled by the introduction of inflation
\cite{guth,rev} which has been shown to be a possible solution of the horizon
and flatness problems.  In addition to model building and other discussions
of how scalar field theories can produce inflation, the question of the
reheating of the universe, the transfer of energy from the inflaton to other
particle modes, has received much attention \cite{reheat}.  The reason for this
is that exponential expansion during inflation causes the universe to
become very cold, as the energy in fields other than the inflaton is redshifted
away.  In order to achieve the standard results of nucleosynthesis, and
possibly other early processes, the universe must be reheated to temperatures
above those at which these important processes take place.  Early efforts to
account for the necessary reheating introduced {\it ad hoc} decay widths to the
inflaton, assuming the energy transfer occurred through single particle decay
\cite{origreheat}.  However, more recently, it has been realized that there are
much more efficient processes, those of either spinodal decomposition in the
case of a (new) inflationary phase transition and 
parametric amplification in the case of chaotic inflationary scenarios
\cite{branden,kls,us1,par,tkachev,son,symrest,kaiser,yoshimura,usfrwprl,kaiser2}. 
Such mechanisms, in which the rate of energy transfer grows 
exponentially, are referred to generally as preheating.

Recently we have treated the dynamical processes of inflation and
preheating\cite{us1} within a fully non-equilibrium
formalism\cite{ctp,hu,ringwald,leutwyler,weiss}.
Such a treatment is necessary due to
the expansion of the universe and the nature of the rapidly evolving dynamics
of preheating.  Non-equilibrium analyses have shown a wealth of new phenomena
which were missed in equilibrium
studies\cite{us1,mink,usfrwprl,frw,usDeS,kaiser2}. 

It has also become clear that the early analyses in which the inflaton is
treated as a classical field or within perturbation theory are, in many cases,
inadequate.  In fact, despite the tiny couplings usually assumed for the
inflaton, the unstable growth of modes during preheating causes the dynamics to
become non-perturbative\cite{largen,jackiwetal,guven}.

In this article, we study the dynamics of scalar fields in an expanding
isotropic universe using the non-equilibrium closed time path formalism (CTP)
\cite{ctp}, keeping track of the evolution of both the zero mode of the
inflaton and its fluctuations.  We treat the dynamics
using two non-perturbative schemes.  These are the Hartree approximation 
appropriate to a theory with discrete symmetry and the leading
order large $N$ approximation of an $O(N)$ vector model which describes
theories with continuous symmetry, satisfying the corresponding Ward identities.
\cite{us1,mink,frw,largenfrw,cooper}.  Both of these are mean field theory
approximations, and as such they cannot account for the particle scattering
processes that would allow the universe to reenter the hot Big Bang scenario
after inflation. Eventually, these processes should be taken into account to
determine the final reheating temperature. Here we will concern ourselves
only with processes occurring before thermalization.

In particular, we will study the process of preheating using a wide range of
initial conditions while the study of the inflationary stage will be discussed
elsewhere \cite{usDeS}.  Taking our cues from the evolution of the equation of
state in the inflationary and preheating phases, we analyze the dynamics of
preheating in {\it fixed} radiation and matter dominated
Friedman-Robertson-Walker cosmologies both analytically and numerically.  We
follow the equation of state during the evolution to ensure that our evolution
obeys the appropriate gravitational dynamics.

In the next section, we set up the closed time path formalism, describe our
models and the two approximations we will use to study the evolution, and write
down our evolution equations for the zero mode and the fluctuations.  In
section III, we discuss some important issues regarding the renormalization
aspects of the problem, including the renormalization of the energy and
pressure densities in an expanding background.  Section IV begins with early
time solutions in the slow roll scenario followed by a full numerical analysis
of the various cases.  We conclude the section with late time analyses of the
large $N$ evolution equations in the case of a symmetry broken potential.  In
the conclusions, we contrast our work with other analyses in the literature and
discuss avenues which may be pursued to further improve our knowledge of these
important cosmological problems.  An appendix is provided with a discussion of
our choice of initial conditions and their physical implications.

Our main results are as follows.  In the situations we analyze, we
find that the expansion of the universe allows for significant particle
production, although this production is somewhat sensitive to the exact
expansion rate and is effectively shut off for high enough rates.  
In the case of a symmetry broken potential, we
determine that in the large $N$ limit, the quantum fluctuations
decrease for  late times as $1/a^2(t)$, while these fluctuations and
the zero mode 
satisfy a sum rule consistent with Goldstone's theorem.  In addition to these
results, we present a consistent renormalization of the energy
momentum tensor in a flat FRW spacetime within both the large $N$ and the
Hartree approximations. Such a result is an essential component of any
consistent analysis of the backreaction problem in an expanding universe. 

We compute the renormalized energy density, $\varepsilon$, and the
pressure, $p$,  as a function of time. Averaging over the field oscillations,
we find immediately after preheating a cold matter equation of state ($ p = 0 $) 
in the slow roll scenarios.  In chaotic scenarios, the equation of state 
just after preheating is between that of radiation ($p = \varepsilon/3$) and
matter where the matter dominated regime is reached only for late times.
The timescale over which the equation of state becomes matter dominated
depends on the distribution of created particles in momentum space in 
addition to the approximation scheme implemented.

\section{The Formalism and Models}

We present here the framework of the non-equilibrium closed time path formalism.
For a more complete discussion, the reader is referred to \cite{us1}, or the 
alternative approach given in \cite{frw}. 

The time evolution of a system is determined in the Schr\"odinger picture
by the functional Liouville equation
\begin{equation}
i\frac{\partial \rho (t)}{\partial t} = [H(t),\rho(t)],
\label{liouville}
\end{equation}
where $\rho$ is the density matrix and we allow for an explicitly time
dependent Hamiltonian as is necessary to treat quantum fields in a time
dependent background.  Formally, the solutions to this equation for the time
evolving density matrix are given by the time evolution operator, $U(t,t^{'})$,
in the form
\begin{equation}
\rho(t) = U(t,t_0)\rho(t_0)U^{-1}(t,t_0).
\label{rhoevol}
\end{equation}
The quantity $\rho(t_0)$ determines the initial condition for the evolution. We
choose this initial condition to describe a state of local equilibrium in
conformal time, which is also identified with the conformal adiabatic vacuum
for short wavelengths.  In the appendix we provide an analysis and discussion
of different initial conditions and their physical content within the context
of expanding cosmologies.

Given the evolution of the density matrix (\ref{rhoevol}), ensemble averages
of operators are given by the expression (again in the Schr\"odinger picture)
\begin{equation}
\langle{\cal O}(t)\rangle = 
\frac{Tr[U(t_0,t){\cal O}U(t,t^{'})U(t^{'},t_0)\rho(t_0)]}{Tr\rho(t_0)},
\label{expect}
\end{equation}
where we have inserted the identity, $U(t,t^{'})U(t^{'},t)$ with $t^{'}$ an
arbitrary time which will be taken to infinity. The state is first evolved
forward from the initial time $t_0$ to $t$ when the operator is inserted.  We
then evolve this state forward to time $t^{'}$ and back again to the initial
time.

The actual evolution of various quantities in the theory can now be evaluated
by either constructing the appropriate Green functions as in \cite{us1}, or by
choosing an explicit ansatz for the functional form of the time dependent
density matrix so that the trace in (\ref{expect}) may be explicitly evaluated
as a functional integral (see \cite{frw}).  The methods are equivalent, and
provide the results which will be presented below for the cases of interest.

Since we are currently interested in the problem of preheating at the end of
inflation, we will work in a spatially flat Friedman-Robertson-Walker
background with scale factor $a(t)$ and line element:
\begin{equation}
ds^2 = dt^2 - a^2(t)\; d\vec{x}^2.
\label{metric}
\end{equation}
Our Lagrangian density has the form
\begin{equation}
{\cal L} =\sqrt{-g}\left[ \frac12 \nabla_\mu\Phi\nabla^\mu\Phi - V(\Phi)\right].
\label{lagrangian}
\end{equation}

\subsection{Hartree Approximation}
In the Hartree approximation, our theory is that of a single component scalar
field, $\Phi(\vec{x},t)$, with the $Z_2$ symmetry $\Phi \to -\Phi$.  The
potential can be written as:
\begin{equation}
V(\Phi) = \frac12 (m^2+\xi{\cal R}) \Phi^2 + \frac{\lambda}{4!}\Phi^4,
\label{hartpot}
\end{equation}
where ${\cal R}$ is the Ricci scalar.
We decompose the field into its zero mode, $\phi(t) =
\langle\Phi(\vec{x},t)\rangle$, and fluctuations $\psi(\vec{x}, t)$ about it:
\begin{equation}
\Phi(\vec{x},t) = \phi(t) + \psi(\vec{x},t).
\label{hartdecomp}
\end{equation}
The potential (\ref{hartpot}) may then be expanded in terms of these fields.

The Hartree approximation is achieved by making the potential quadratic in
the fluctuation field $\psi$ by invoking the factorization
\begin{eqnarray}
\psi^3(\vec{x},t) & \to & 3\langle\psi^2(\vec{x},t)\rangle\psi(\vec{x},t), 
\label{hartfactor1} \\
\psi^4(\vec{x},t) & \to & 6\langle\psi^2(\vec{x},t)\rangle\psi^2(\vec{x},t)
	-3\langle\psi^2(\vec{x},t)\rangle^2.
\label{hartfactor2}
\end{eqnarray}
This factorization yields a quadratic theory in which the effects of
interactions are encoded in the time dependent mass which is determined
self-consistently.

The equations of motion for the zero mode and the fluctuations are
given by the  tadpole equation
\begin{equation}
\langle\psi(\vec{x},t)\rangle = 0.
\label{tadpole}
\end{equation}
Introducing the Fourier mode functions, $U_k(t)$, they can be written as:
\begin{eqnarray}
\ddot{\phi}(t)+3\frac{\dot{a}(t)}{a(t)}\; \dot{\phi}(t)+(m^2+\xi{\cal
R}(t))\;\phi(t) 
+\frac{\lambda}{6}\;\phi^3(t)+\frac{\lambda}{2}\;\phi(t)
\;\langle\psi^2(t)\rangle 
& = & 0\; ,
\label{hartphieq} \\  \nonumber \\
\left[\frac{d^2}{dt^2}+3\frac{\dot{a}(t)}{a(t)}\frac{d}{dt}+\frac{k^2}{a^2(t)}
+m^2+\xi{\cal R}(t)+\frac{\lambda}{2}\; \phi^2(t)+
\frac{\lambda}{2}\langle\psi^2(t)\rangle\right]
U_k(t) & = & 0\; ,
\label{hartukeq}
\end{eqnarray}
\begin{eqnarray}
\langle\psi^2(t)\rangle &=& \int \frac{d^3k}{2(2\pi)^3}\, |U_k(t)|^2
\; .
\label{fluct} \\ \nonumber
\end{eqnarray}
The initial conditions on the mode functions are
\begin{equation}
U_k(t_0) = \frac{1}{\sqrt{\omega_k(t_0)}}, \quad \dot{U}_k(t_0) =\left[
-\frac{\dot{a}(t_0)}{a(t_0)}-i\omega_k(t_0)\right]U_k(t_0),
\label{initcond}
\end{equation}
with the frequencies $\omega_k(t_0)$ given by
\begin{equation}
\omega_k(t_0) = \left[k^2+{\cal{M}}^2(t_0)\right]^{1/2} \; \; , \; \;
{\cal{M}}^2(t) = a^2(t)\left[m^2+(\xi-1/6){\cal R}(t)+
\frac{\lambda}{2}\phi^2(t)+\frac{\lambda}{2}\langle\psi^2(t)\rangle \right].
\label{freq}
\end{equation}
A detailed analysis and discussion of the choice of initial conditions and the
frequencies (\ref{freq}) is provided in the appendix. As discussed there, this
choice corresponds to the large-$k$ modes being in the conformal adiabatic
vacuum state. In what follows we will subtract the composite operator
$\psi^2(t)$ at the initial time and absorb the term
$\frac{\lambda}{2}\langle\psi^2(t_0)\rangle$ in a renormalization of the mass.
Furthermore, we choose the scale of time such that $a(t_0)=1$ in both radiation 
and matter dominated cosmologies.

\subsection{Large $N$ Limit}
To discuss the large $N$ limit, we now treat $\Phi$ as an $N$-component vector
in a theory with the continuous $O(N)$ symmetry.  The potential is
\begin{equation}
V(\Phi) = \frac12 (m^2+\xi {\cal R})\Phi\cdot\Phi + 
\frac{\lambda}{8N}(\Phi\cdot\Phi)^2.
\label{Npot}
\end{equation}
We now break up the field $\Phi$ 
into a single scalar field $\sigma$ and an $N-1$ component vector
$\vec{\pi}$ as 
$\Phi = (\sigma,\vec{\pi})$ and allow the $\sigma$ field to have a non-zero
expectation value.  Taking the decomposition 
\begin{equation}
\sigma(\vec{x},t)= \sqrt{N} \phi(t) + \chi(\vec{x},t),
\label{Ndecomp}
\end{equation}
where $\sqrt{N} \phi(t)$ is the expectation value of $\sigma$ and $\chi$ is the
fluctuation about this value. If we write the ``pion'' field as
\begin{equation}
\vec{\pi}(\vec{x},t) = \psi(\vec{x},t)\overbrace{(1,1,\ldots,1)}^{N-1},
\label{pions}
\end{equation}
we reach the leading order large $N$ limit by assuming the factorization
\begin{equation}
\psi^4(\vec{x},t) \to 2\langle\psi^2(\vec{x},t)\rangle\psi^2(\vec{x},t)
-\langle\psi^4(\vec{x},t)\rangle,
\label{Nfactor}
\end{equation}
with
\begin{equation}
\langle\psi^2\rangle = O(1), \quad 
\langle\chi^2\rangle = O(1), \quad
\phi = O(1).
\label{Nreq}
\end{equation}
We see that since there are $N-1$ pion fields, contributions from the field
$\chi$ can be neglected in the formal limit as they are of order $1/N$ with
respect those of $\psi$ and $\phi$.

Again, we determine the equations of motion via the condition (\ref{tadpole}).
The zero mode equation becomes
\begin{equation}
\ddot{\phi}(t)+3\frac{\dot{a}(t)}{a(t)}\dot{\phi}(t)+[m^2+\xi{\cal
R}(t)]\phi(t) 
+\frac{\lambda}{2}\phi^3(t)+\frac{\lambda}{2}\; \phi(t)\;
\langle\psi^2(t)\rangle =0,  \label{Nphieq}
\end{equation}
while the equations for the modes, the fluctuation, and the initial conditions
are identical to the Hartree case given by equations (\ref{hartukeq}) -
(\ref{initcond}).  Notice that we have used identical notations in the two
cases to avoid cluttering and also to stress the similarity between the two
approximations.  In particular, we note that the only difference in the
expressions for the two cases [eqs.(\ref{hartukeq}) and (\ref{Nphieq}),
respectively] is a factor of three appearing in the self interaction term in
the equations for the zero mode.  However, as we will see, the two approximations
are describing theories with distinct symmetries and there will be qualitative 
differences in the results.  

An important point to note in the large $N$ equations of motion is that the
form of the equation for the zero mode (\ref{Nphieq}) is the same as for the
$k=0$ mode function (\ref{hartukeq}). It will be this identity that allows
solutions of these equations in a symmetry broken scenario to satisfy
Goldstone's theorem.

\section{Renormalization and Energy Momentum}

Upon examination of the equal time correlator (\ref{fluct}), one finds that the
integral is divergent and thus must be regulated.  This can be done by a number
of methods, but we require a scheme which is amenable to numerical calculation.
We therefore introduce a large momentum cutoff which renders the integral
finite, and one finds that it is possible to remove the terms depending both
quadratically and logarithmically on the cutoff by a renormalization of the
parameters of the theory\cite{frw}. After these terms are subtracted, the
cutoff may be taken to infinity, and the remaining quantity is both physical
and finite.  A similar process is required to regulate the expressions for the
energy density and the pressure as will be described in more detail below
\cite{usDeS}.

In terms of the variables introduced in section II above, the renormalization
of (\ref{fluct}) proceeds almost identically in the Hartree and large $N$
approximations.  The WKB analysis that reveals the large-$k$ behavior of the mode
functions is described in detail in the appendix wherein we quote the relevant
expressions for the large-$k$ behavior for the mode functions and their
derivatives.  Denoting bare and renormalized parameters with subscripts $b$ and
$r$ respectively, and the momentum cutoff and subtraction point as $\Lambda$
and $\kappa$ respectively, and using the results of the appendix, we find the
following renormalization scheme:
\begin{eqnarray}
m_b^2 & + & \frac{\lambda_b}{16\pi^2}\frac{\Lambda^2}{a^2(t)} = 
m_r^2\left[1+\frac{\lambda_b}{16\pi^2}\ln(\Lambda/\kappa)\right],
\label{massrenorm} \\
\lambda_b & = & \frac{\lambda_r}{1-\gamma\lambda_r\ln(\Lambda/\kappa)/16\pi^2},
\label{couprenorm} \\
\xi_b & = & \xi_r + \frac{\lambda_b}{16\pi^2}(\xi_r-1/6)\ln(\Lambda/\kappa),
\label{xirenorm}
\end{eqnarray}
where $\gamma$ is a combinatorial factor taking on the value $\gamma=1$ in the
large $N$ limit and $\gamma=1/3$ in the Hartree approximation.  The subtracted
equal time correlator is now given by
\begin{eqnarray}
\langle\psi^2(t)\rangle_r & = & \int^{\Lambda}\frac{d^3k}{(2\pi)^3}\left\{
\frac{|U_k(t)|^2}{2} 
- \frac{1}{2ka^2(t)}\right. \nonumber \\
& + & \left.\frac{\theta(k-\kappa)}{4k^3}\left[(\xi_r-1/6){\cal R}(t)+m_r^2+
\frac{\lambda_r}{2}(\phi^2(t)+\langle\psi^2(t)\rangle_r)\right]\right\}.
\label{fluctrenorm}
\end{eqnarray}

A further finite subtraction at $t_0$ is performed and absorbed in a further
finite and time independent renormalization of the mass.  Notice in particular
that in order for the renormalization of the mass to be time independent, we
must require that the cutoff $\Lambda$ be fixed in {\it physical} coordinates
and therefore have the form $\Lambda \propto a(t)$.

Our treatment of the renormalization of the energy momentum tensor is similar
to the approach of \cite{anderson}, extended to the non-perturbative Hartree
and large $N$ approximations.  The expressions for the expectation values of
the energy density, $\varepsilon$, and the trace of the stress energy,
$\varepsilon-3p$, where $p$ is the pressure density are:
\begin{eqnarray}
\frac{\varepsilon}{N} & = & \frac12\dot{\phi}^2 + \frac12m^2\phi^2 +
\frac{\gamma\lambda}{8}\phi^4 + \frac{m^4}{2\gamma\lambda} - \xi
G^{0}_{0}\phi^2 +  6\xi\frac{\dot{a}}{a}\phi\dot{\phi}
\nonumber \\
& + & \frac12\langle\dot{\psi}^2\rangle + 
\frac{1}{2a^2}\langle(\nabla\psi)^2\rangle + \frac12 m^2 \langle\psi^2\rangle 
+ \frac{\lambda}{8}[2\phi^2\langle\psi^2\rangle + \langle\psi^2\rangle^2] 
\nonumber \\
& - & \xi G^{0}_{0}\langle\psi^2\rangle + 
6\xi\frac{\dot{a}}{a}\langle\psi\dot{\psi}\rangle, \label{energy} \\
\frac{\varepsilon-3p}{N} & = & -\dot{\phi}^2 + 2m^2\phi^2 + 
\frac{\gamma\lambda}{2}\phi^4 + \frac{2m^4}{\gamma\lambda} - \xi
G^{\mu}_{\mu}\phi^2 + 
6\xi\left(\phi\ddot{\phi} + \dot{\phi}^2 +
3\frac{\dot{a}}{a}\phi\dot{\phi}\right) 
\nonumber \\
& - & (1-6\xi)\langle\dot{\psi}^2\rangle +
\frac{1-6\xi}{a^2}\langle(\nabla\psi)^2\rangle + 
(2-6\xi)m^2\langle\psi^2\rangle - \xi
G^{\mu}_{\mu}(1-6\xi)\langle\psi^2\rangle  
\nonumber \\
& + & \frac{\lambda}{2}[(2-6\xi)\phi^2\langle\psi^2\rangle +
\langle\psi^2\rangle^2 - 
6\xi\langle\psi^2\rangle\langle\psi^2\rangle_r],
\label{trace}
\end{eqnarray}
where we have used the equations of motion in deriving this expression for the
trace (\ref{trace}). The quantities $G^{\mu}_{\mu}=-{\cal R}$ and $
G^{0}_{0}=-3(\dot{a}/a)^2 $ are the trace and the time-time components of the
Einstein curvature tensor, $\langle\psi^2\rangle$ is given by equation
(\ref{fluct}), $\langle\psi^2\rangle_r$ by (\ref{fluctrenorm}), and we have
defined the following integrals:
\begin{eqnarray}
\langle(\nabla\psi)^2\rangle & = & \int \frac{d^3k}{2 (2\pi)^3}k^2
|U_k(t)|^2 \; , \label{delpsi} \\
\langle\dot{\psi}^2\rangle & = & \int \frac{d^3k}{2(2\pi)^3}
|\dot{U}_k(t)|^2 \;. \label{dotpsi}
\end{eqnarray}
The composite operator $\langle \psi \dot{\psi} \rangle$ is symmetrized by
removing a normal ordering constant to yield
\begin{equation}
\frac{1}{2}(\langle\psi\dot{\psi}\rangle + \langle\dot{\psi}\psi\rangle)
 =  \frac{1}{4} \int \frac{d^3k}{(2\pi)^3}
\frac{d |U_k(t)|^2}{dt}. 
\label{psidotpsi} 
\end{equation}
Each of these integrals is divergent and must be regularized.  We proceed in
the same manner as above, imposing an ultraviolet cutoff, $\Lambda$, and
computing the high $k$ expansions of the $U_k$, this time to fourth order in
$1/k$.  We find the following divergences in $\varepsilon$ and
$\varepsilon-3p$:
\begin{eqnarray}
\left(\frac{\varepsilon}{N}\right)_{div} & = & \frac{\Lambda^4}{16\pi^2a^4} +
\frac{\Lambda^2}{16\pi^2a^2}\left[2(\xi_r-1/6)G^{0}_{0}+m_r^2+
\frac{\lambda_r}{2}(\phi^2+\langle\psi^2\rangle_r)\right] \nonumber \\
& + & \frac{\ln(\Lambda/\kappa)}{16\pi^2}\left[-\frac{m_r^4}{2} -
m_r^2\frac{\lambda_r}{2}(\phi^2+\langle\psi^2\rangle_r) -
\frac{\lambda_r^2}{8}(\phi^2+\langle\psi^2\rangle_r)^2\right. \nonumber \\
& + & \left. 2(\xi_r-1/6)G^{0}_{0}\left(m_r^2+
\frac{\lambda_r}{2}(\phi^2+\langle\psi^2\rangle_r)\right) 
+(\xi_r-1/6)^2H^{0}_{0}\right. \nonumber \\
& - & \left.6(\xi_r-1/6)\frac{\dot{a}}{a}\frac{\lambda_r}{2}
\frac{d}{dt}(\phi^2+\langle\psi^2\rangle_r)\right], 
\label{energydiv} \\
\left(\frac{\varepsilon-3p}{N}\right)_{div} & = & \frac{\Lambda^2}{16\pi^2a^2}
\left[2(\xi_r-1/6)G^{\mu}_{\mu} 
+ 12(\xi_r-1/6)\frac{\dot{a}^2}{a^2} + 2m_r^2 + 
\lambda_r(\phi^2+\langle\psi^2\rangle_r)\right] \nonumber \\
& + & \frac{\ln(\Lambda/\kappa)}{16\pi^2}\left[-2m_r^4 - 
2m_r^2\lambda_r(\phi^2+\langle\psi^2\rangle_r) - \frac{\lambda_r^2}{2}
(\phi^2+\langle\psi^2\rangle_r)^2\right. \nonumber \\
& + & \left. 2(\xi_r-1/6)G^{\mu}_{\mu}
\left(m_r^2+\frac{\lambda_r}{2}(\phi^2+\langle\psi^2\rangle_r)\right)\right. 
+ (\xi_r-1/6)^2H^{\mu}_{\mu}
\nonumber \\ 
& - & \left. 6(\xi_r-1/6)\left[\frac{\lambda_r}{2}\frac{d^2}{dt^2}
(\phi^2+\langle\psi^2\rangle_r) +
3\frac{\dot{a}}{a}\frac{\lambda_r}{2}\frac{d}{dt}
(\phi^2+\langle\psi^2\rangle_r)\right]\right].
\label{tracediv}
\end{eqnarray}

The quantities $H^{0}_{0}$ and $H^{\mu}_{\mu}$ are the time-time component and
trace of a geometrical tensor given by the variation with respect to the metric
of higher derivative terms appearing in the gravitational action (such as
${\cal R}^2$).  For the present case, they are given in terms of the Ricci
scalar by the expressions
\begin{eqnarray}
H^{0}_{0} & = & -6\left(\frac{\dot{a}}{a}\dot{{\cal R}} + 
\frac{\dot{a}^2}{a^2}{\cal R} - \frac{1}{12}{\cal R}^2\right), 
\label{hzerozero} \\
H^{\mu}_{\mu} & = & -6\left(\ddot{{\cal R}} + 
3\frac{\dot{a}}{a}\dot{{\cal R}}\right).
\label{htrace}
\end{eqnarray}

Eqs.(\ref{energydiv}) and (\ref{tracediv}) are closely related to eqs.(3.17) in
ref.\cite{usDeS}. However, they are not identical since the initial conditions
chosen in ref.\cite{usDeS} were different from the ones selected in the present
paper. Furthermore, we had $ \xi_r = 0 $ in ref.\cite{usDeS}.

The energy momentum is made finite by subtraction of the divergent pieces
(\ref{energydiv}) and (\ref{tracediv}) from the expressions for the energy
density (\ref{energy}) and the trace (\ref{trace}).  Within the context of
covariant regularization schemes such as dimensional regularization and
covariant point splitting\cite{birrell}, such a procedure has been shown to
adequately renormalize couplings appearing in the semi-classical Einstein's
equation, extended to account for higher derivative terms and a possible
cosmological constant, $K$.  This equation has the form
\begin{equation}
\frac{G^{\mu}_{\nu}}{8\pi G_N} + \alpha H^{\mu}_{\nu} + 
\frac{K g^{\mu}_{\nu}}{8\pi G_N} = - \langle T^{\mu}_{\nu} \rangle,
\end{equation}
where $G_N$ is Newton's constant and $\alpha$ is the coupling to the higher
order gravitational term.  However, regularization via an ultraviolet cutoff is
not a covariant scheme and we find that the quadratic and quartic divergence
structure of the energy momentum in this scheme does not have the correct form
to consistently renormalize the parameters of the semi-classical theory.
Nevertheless, the subtraction procedure described here produces a stress energy
tensor which is both {\it finite} and {\it covariantly conserved} in addition
to being amenable to numerical study.  An alternative renormalized
computational scheme in which covariant regularization is possible is presented
by Baacke, Heitmann, and P\"atzold \cite{baacke}.  It should be possible to
extend such a scheme to expanding spacetimes.

\section{Evolution}

We focus our study of the evolution on radiation or matter dominated
cosmologies, as the case for de Sitter expansion has been studied
previously\cite{usDeS}.  We write the scalefactor as $a(t)=(t/t_0)^n$ with
$n=1/2$ and $n=2/3$ corresponding to radiation and matter dominated
backgrounds respectively.  Note that the value of $t_0$ determines the
initial Hubble constant since 
$$
H(t_0)=\frac{\dot{a}(t_0)
}{a(t_0)}=\frac{n}{t_0}.
$$
We now solve the system of equations (\ref{hartphieq}) - (\ref{fluct}) in the
Hartree approximation, with (\ref{Nphieq}) replacing (\ref{hartphieq}) in the
large $N$ limit.  We begin by presenting an early time analysis of the
slow roll scenario.  We then undertake a thorough numerical investigation of
various cases of interest.  For the symmetry broken case, we also provide an
investigation of the late time behavior of the zero mode and the quantum
fluctuations.

In what follows, we scale all variables in terms of the magnitude of the
renormalized mass, taking $|m_r^2|=1$.  We also define the variable
$$
\eta^2(t)\equiv\lambda\phi^2(t)/2
$$ 
and write 
$$
g\Sigma(t)\equiv\lambda\langle\psi^2(t)\rangle_r/2
\; , \; g \equiv \lambda/8\pi^2 = 10^{-12} \; .
$$
We drop the subscript $r$
denoting the renormalized parameters, and we will assume minimal coupling to
the curvature, $\xi_r = 0$.  In the most of the cases of interest, 
${\cal R} \ll 1$, so that finite $\xi_r$ has little effect.

\subsection{Early Time Solutions for Slow Roll}

For early times in a slow roll scenario [$m^2=-1$, $\eta(t_0) \ll 1$], we
can neglect in eqs.(\ref{hartphieq}) or (\ref{Nphieq}) and in
eq.(\ref{hartukeq}) both the quadratic and cubic terms in $\eta(t)$ as well as
the quantum fluctuations $\langle\psi^2(t)\rangle_r $ [recall that
$\langle\psi^2(t_0)\rangle_r = 0 $]. Thus, the differential equations for the
zero mode (\ref{hartphieq}) or (\ref{Nphieq}) and the mode functions
(\ref{hartukeq}) become linear equations. In terms of the scaled variables
introduced above, with $ a(t)= (t/t_0)^n $ ($ n=2/3 $ for a matter dominated
cosmology while $ n=1/2 $ for a radiation dominated cosmology) we have:
\begin{eqnarray}
\ddot{\eta}(t)+\frac{3n}{t}\dot{\eta}(t)-\eta(t) & = & 0 \; ,
\label{earlyeta} \\ \nonumber \\
\left[\frac{d^2}{dt^2}+\frac{3n}{t}\frac{d}{dt}
+\frac{k^2}{(t/t_0)^{2n}}-1\right]U_k(t) & = & 0 \;. \label{earlyuk}
\end{eqnarray}

The solutions to the zero mode equation (\ref{earlyeta}) are
\begin{equation}
\eta(t)=c\; t^{-\nu}I_{\nu}(t)+d\; t^{-\nu}K_{\nu}(t) \; ,
\label{earlyetasoln}
\end{equation}
where $\nu \equiv (3n-1)/2$, and $I_{\nu}(t)$ and $K_{\nu}(t)$ are modified
Bessel functions.  The coefficients, $c$ and $d$, are determined by the initial
conditions on $\eta$.  For $\eta(t_0)=\eta_0$ and $\dot{\eta}(t_0)=0$, we have:
\begin{eqnarray}
c & = & \eta_0  \; t_0^{\nu+1}
\left[ \dot{K}_{\nu}(t_0)-\frac{\nu}{t_0}K_{\nu}(t_0)\right] \; ,
\label{coeffc} \\
d & = & -\eta_0  \; t_0^{\nu+1}
\left[\dot{I}_{\nu}(t_0)-\frac{\nu}{t_0}I_{\nu}(t_0)\right] \; .
\label{coeffd}
\end{eqnarray}
Taking the asymptotic forms of the modified Bessel functions, we find that for
intermediate times $\eta(t)$ grows as
\begin{equation}
\eta(t) \stackrel{t \gg 1}{=} {c \over {\sqrt{2 \pi}}}
\; t^{-3n/2}\; e^t\left[1-\frac{9n^2-6n}{8t}+
O\left({1\over{t^2}}\right)\right].
\label{asymeta}
\end{equation}
We see that $\eta(t)$ grows very quickly in time, and the approximations
(\ref{earlyeta}) and (\ref{earlyuk}) will quickly break down.  For the case
shown in figure 1 (with $n=2/3$, $H(t_0)=0.1$, $\eta(t_0)=10^{-7}$, 
and $\dot{\eta}(t_0)=0$),
we find that this approximation is valid up to $t-t_0 \simeq 10$.

The equations for the mode functions ({\ref{earlyuk}) can be solved in closed
form for the modes in the case of a radiation dominated cosmology with $n=1/2$.
The solutions are

\begin{equation}
U_k(t) = c_k \; e^{-t}\, U\left(\frac34-\frac{k^2 t_0^{2n}}{2},\frac32,2t\right) 
+ d_k \; e^{-t}\, M\left(\frac34-\frac{k^2 t_0^{2n}}{2},\frac32,2t\right).
\label{earlyuksoln}
\end{equation}
Here, $U(\cdot)$ and $M(\cdot)$ are confluent hypergeometric functions
\cite{aands} (in another common notation, $M(\cdot) \equiv\; _1F_1(\cdot)$),
and the $c_k$ and $d_k$ are coefficients determined by the initial conditions
(\ref{initcond}) on the modes.  The solutions can also be written in terms of
parabolic cylinder functions.

For large $ t $ we have the asymptotic form
\begin{eqnarray}
U_k(t) & \stackrel{t \gg 1}{=} & d_k \; e^t (2t)^{-(3/4+(k t_0^n)^2/2)}
\frac{\sqrt{\pi}}{2\;\Gamma\left(\frac34-\frac{(k t_0^n)^2}{2}\right)}
\left[1+O\left({1\over{t}}\right)\right] \\ \nonumber
& + & c_k \; e^{-t}\,  (2t)^{(-3/4+(k t_0^n)^2/2)}\left[1+O
\left({1\over{t}}\right)\right]\;
. \label{asymuk}
\end{eqnarray}
Again, these expressions only apply for intermediate times before the
nonlinearities have grown significantly.

\subsection{Numerical Analysis}
We now present the numerical analysis of the dynamical evolution of
scalar fields in time dependent, matter and radiation dominated cosmological
backgrounds.  We use initial values of the Hubble constant such that 
$H(t_0) \geq 0.1$.  For expansion
rates much less than this value the evolution will look similar to
Minkowski space, which has been studied in great detail elsewhere
\cite{us1,mink,cooper}.  As will be seen, the equation of state found
numerically is, in the majority of cases, that of cold matter.  We therefore
use matter dominated expansion for the evolution in much of the analysis that
follows.  While it presents some inconsistency at late times, 
the evolution in radiation
dominated universes remains largely unchanged, although there is greater
initial growth of quantum fluctuations due to the scale factor growing more
slowly in time.  Using the large $N$ and Hartree approximations to study 
theories with continuous and discrete symmetries respectively, we treat three
important cases.  They are 1) $m^2<0$, $\eta(t_0)\ll 1$; 2) $m^2<0$,
$\eta(t_0)\gg 1$; 3) $m^2>0$, $\eta(t_0)\gg 1$.

In presenting the figures, we have shifted the origin of time such that
$t \to t'=t-t_0$.  This places the initial time, $t_0$, at the origin.
In these shifted coordinates, the scalefactor is given by 
$$
a(t)=\left(\frac{t+\tau}{\tau}\right)^n,
$$ 
where, once again, $n=2/3$ and $n=1/2$ in matter and radiation dominated
backgrounds respectively, and the value of $\tau$ is determined by the 
Hubble constant at the initial time:
$$
H(t_0=0)=\frac{n}{\tau}.
$$

{\bf Case 1: $m^2<0$, $\eta(t_0)\ll 1$}.  This is the case of an early universe
phase transition in which there is little or no biasing in the initial
configuration (by biasing we mean that the initial conditions break the $ \eta
\to -\eta $ symmetry).  The transition occurs from an initial temperature above
the critical temperature, $T>T_c$, which is quenched at $t_0$ to the
temperature $T_f \ll T_c$.  This change in temperature due to the rapid
expansion of the universe is modeled here by an instantaneous change in the
mass from an initial value $m_i^2=T^2/T_c^2-1$ to a final value
$m_f^2=-1$.  We will use the value $m_i^2=1$ in what follows.
This quench approximation is necessary since the low momentum
frequencies (\ref{freq}) appearing in our initial conditions (\ref{initcond}) 
are complex for negative mass squared and small $\eta(t_0)$.  An alternative
choice is to use initial frequencies given by
$$
\omega_k(t_0)=\left[k^2+{\cal{M}}^2(t_0)\tanh\left(
\frac{k^2+{\cal{M}}^2(t_0)}{|{\cal{M}}^2(t_0)|}\right)\right]^{1/2}.
$$
These frequencies have the attractive feature that they match the conformal
adiabatic frequencies given by (\ref{freq}) for large values of $k$ while 
remaining positive for small $k$.  We find that such a choice of initial
conditions changes the quantitative value of the particle number by a few 
percent, but leaves the qualitative results unchanged.

While this case should show some qualitative features of the corresponding
process of preheating in slow-roll inflation, we note that since quantum
fluctuations can grow to be large during the de Sitter phase (see
\cite{usDeS}), a proper treatment of preheating after slow-roll inflation must
account for the full gravitational backreaction.  This will be the subject of a
future article \cite{grav}.  In the present case, we impose a background
cosmology dominated by ordinary radiation or matter. 
We plot the the zero mode $\eta(t)$, the equal time correlator $g\Sigma(t)$, 
the total number of produced particles $gN(t)$  (see the appendix for
a discussion of our definition of particles), the number of particles
$gN_k(t)$ as a function of wavenumber for both intermediate and
late times, and the ratio of the pressure
and energy densities $p(t)/\varepsilon(t)$ (giving the equation of state).

Figure 1a-e shows these quantities in the large $N$ approximation for a matter
dominated cosmology with an initial condition on the zero mode given by
$\eta(t_0\! =\! 0)=10^{-7}$, $\dot{\eta}(t_0\! =\! 0)=0$ and for an initial
expansion rate of $H(t_0)=0.1$.  This choice for
the initial value of $\eta$ stems from the fact that the quantum fluctuations
only have time to grow significantly for initial values satisfying
$\eta(t_0) \ll  \sqrt{g}$; for values $\eta(t_0) \gg \sqrt{g}$ the evolution
is essentially classical.  This result is clear from the intermediate time
dependence of the zero mode and the low momentum mode functions given by 
the expressions (\ref{asymeta}) and (\ref{asymuk}) respectively.

After the initial
growth of the fluctuation $g\Sigma$ (fig. 1b) we see that the zero mode
(fig. 1a) approaches the value given by the minimum of the tree level
potential, $\eta=1$, while $g\Sigma$ decays for late times as
$$
g\Sigma \simeq {{\cal C}\over{a^2(t)}} ={{\cal C}\over{t^{4/3}}} \; .
$$
For these late times, the Ward identity corresponding to the $O(N)$ symmetry of
the field theory is satisfied, enforcing the condition
\begin{equation}
-1 + \eta^2(t) + g\Sigma(t) = 0.
\label{ward}
\end{equation} 
Hence, the zero mode approaches the classical minimum as
$$
\eta^2(t) \simeq 1 -{{\cal C}\over{a^2(t)}} \; .
$$ 

Figure 1c depicts the number of particles produced.  After an initial burst of 
particle production, the number of particles settles down to a relatively
constant value.  Notice that the number of particles produced is approximately
of order $1/g$.
In figure 1d, we show the number of particles as a function of the
wavenumber, $k$.  For intermediate times we see the simple structure depicted
by the dashed line in the figure, while for late times this quantity becomes
concentrated more at low values of the momentum $k$.  

Finally, figure 1e shows that the field begins with a de Sitter equation of 
state $p=-\varepsilon$ but evolves quickly to a state dominated by ordinary matter, 
with an equation of state (averaged over the oscillation timescale) $p=0$.
This last result is a bit surprising as one expects from the condition
(\ref{ward}) that the particles produced in the final state are massless
Goldstone bosons (pions) which should have the equation of state of radiation.
However, as shown in figure 1d, the produced particles are of low momentum,
$k \ll 1$, and while the effective mass of the particles is zero to very high 
accuracy when averaged over the oscillation timescale, the effective mass 
makes small oscillations about zero so that the dispersion relation for these
particles differs from that of radiation.  In addition, since the produced
particles have little energy, the contribution to the energy density from
the zero mode, which contributes to a cold matter equation of state, remains
significant.

In figures 2a-e we show the same situation depicted in figure 1 using the
Hartree approximation.  The initial condition on the zero mode is $\eta(t_0\!
=\! 0)=\sqrt{3}\cdot 10^{-7}$; the factor of $\sqrt{3}$ appears due to the
different scaling in the zero mode equations, (\ref{hartphieq}) and
(\ref{Nphieq}), which causes the minimum of the tree level effective potential
in the Hartree approximation to have a value of $\eta=\sqrt{3}$.  Again, the
Hubble constant has the value $H(t_0)=0.1$.  Here, we see
again that there is an initial burst of particle production as $g\Sigma$
(fig. 2b) grows large.  However, the zero mode (fig. 2a) quickly reaches the
minimum of the potential and the condition
\begin{equation}
-1 + \eta^2(t)/3 + g\Sigma(t) = 0
\label{ward2}
\end{equation}
is approximately satisfied by forcing the value of $g\Sigma$ quickly to zero.
There are somewhat fewer particles produced here compared to the large $N$ case, 
and the distribution of particles is more extended.  
Since the effective mass of the particles is nonzero, we expect a matter dominated
equation of state (fig 2e) for later times.  
The fact that the Hartree approximation does not
satisfy Goldstone's theorem means that the resulting particles must be massive, 
explaining why somewhat fewer particles are produced.

Finally, we show the special case in which
there is no initial biasing in the field, $\eta(t_0\! =\! 0)=0$, 
$\dot{\eta}(t_0\! =\! 0)=0$, and $H(t_0)=0.1$ in figures
3a-d.  With such an initial condition, the Hartree approximation and the 
large $N$ limit are equivalent.
The zero mode remains zero for all time, so that the quantity
$ g\Sigma(t) $ (fig. 3a) satisfies the sum rule (\ref{ward}) by reaching 
the value one without decaying for late times. 
Notice that many more particles are produced in this case (fig 3b); the growth
of the particle number for late times is due to the expansion of the universe.  
The particle distribution (fig. 3c) is similar to that of the slow roll case in 
figure 1.  The equation of state (fig. 3d) is likewise similar.

In each of these cases of slow roll dynamics, increasing the Hubble constant
has the effect of slowing the growth of both $\eta$ and $g\Sigma$.  The equation
of state will be that of a de Sitter universe for a longer period before
moving to a matter dominated equation of state.  Otherwise, the dynamics is
much the same as in figs. 1-3.

{\bf Case 2: $m^2<0$, $\eta(t_0)\gg 1$}.  We now examine the case of preheating
occurring in a chaotic inflationary scenario with a symmetry broken potential.
In chaotic inflation, the zero mode begins with a value $\eta(t)\gg 1$.  During
the de Sitter phase, $H \gg 1$, and the field initially evolves classically,
dominated by the first order derivative term appearing in the zero mode
equation (see (\ref{hartphieq}) and (\ref{Nphieq})).  Eventually, the zero mode
rolls down the potential, ending the de Sitter phase and beginning the
preheating phase.  We consider the field dynamics in the FRW universe
where preheating occurs after  the end of inflation. 
We thus take the initial temperature to be zero, $T=0$. 

Figure 4 shows our results for the quantities, $\eta(t)$, $g\Sigma(t)$,
$gN(t)$, $gN_k(t)$, and $p(t)/\varepsilon(t)$ for the evolution in
the large $N$ approximation within a {\em radiation} dominated gravitational
background with $H(t_0)=0.1$.  The initial condition on the zero mode is chosen 
to have the representative value $\eta(t_0\!=\! 0)=4$ with 
$\dot{\eta}(t_0\! =\! 0)=0$.  
Initial values of the zero mode much smaller than this will not produce
significant growth of quantum fluctuations; initial values larger than this
produces qualitatively similar results, although the resulting number of 
particles will be greater and the time it takes for the zero mode to settle
into its asymptotic state will be longer. 

We see from figure 4a that the zero
mode oscillates rapidly, while the amplitude of the oscillation decreases due
to the expansion of the universe.  This oscillation induces particle production
through the process of parametric amplification (fig. 4c) and causes the
fluctuation $g\Sigma$ to grow (fig. 4b).  Eventually, the zero mode loses
enough energy that it is restricted to one of the two minima of the tree level
effective potential.  The subsequent evolution closely follows that of Case 1
above with $g\Sigma$ decaying in time as $1/a^2(t) \sim 1/t$ with $\eta$ given by
the sum rule (\ref{ward}).  The spectrum (fig. 4d) indicates a single unstable
band of particle production dominated by the modes $k=1/2$ to about $k=3$ 
for late times.  The structure within this band becomes more complex with time 
and shifts somewhat toward lower momentum modes.  Such a shift has already been 
observed in Minkowski spacetimes \cite{us1}. Figure 4e shows the equation of
state which we see to be somewhere between the relations for matter and
radiation for times out as far as $t=400$, but slowly moving to a matter
equation of state.  Since matter redshifts as $1/a^3(t)$ while radiation
redshifts as $1/a^4(t)$, the equation of state should eventually become matter
dominated.  Given the equation of state indicated by fig. 4e, we estimate that
this occurs for times of order $t=10^4$.  The reason the equation of state
in this case differs from that of cold matter as was seen in figs. 1-3 is 
that the particle distribution produced by parametric amplification is 
concentrated at higher momenta, $k \simeq 1$. 

Figure 5 shows the corresponding case with a matter dominated background.  The
results are qualitatively very similar to those described for figure 4 above.
Due to the faster expansion, the zero mode (fig. 5a) finds one of the two wells
more quickly and slightly less particles are produced.  For late times, the
fluctuation $g\Sigma$ (fig. 5b) decays as $1/a^2(t) \propto 1/t^{4/3}$.  Again we 
see an equation of state (figs. 5e) which evolves from a state between that of
pure radiation or matter toward one of cold matter.

The Hartree case is depicted in figure 6 for a matter dominated universe, with
the initial condition on the zero mode $\eta(t_0\! =\! 0)=4\sqrt{3}$.  Again,
the evolution begins in much the same manner as in the large $N$ approximation
with oscillation of the zero mode (fig. 6a), which eventually settles into one
of the two minima of the effective potential.  Whereas in the large $N$
approximation, the zero mode approaches the minimum asymptotically [as given by
(\ref{ward}) and our late time analysis below], in the Hartree approximation we
see that the zero mode finds the minimum quickly and proceeds to oscillate
about that value.  The two point correlator (fig. 6b) quickly evolves toward
zero without growing large.  Particle production in the Hartree approximation
(figs. 6c-d) is again seen to be inefficient compared to that of the large $N$ 
case above.  Fig. 6e again shows that the equation of state is matter dominated 
for all but the earliest times.

A larger Hubble constant prevents significant particle production unless the 
initial amplitude of the zero mode is likewise increased such that the relation
$\eta(t_0) \gg H(t_0)$ is satisfied.  For very large amplitude $\eta(t_0) \gg 1$, 
to the extent that the mass term can be neglected and while the quantum fluctuation 
term has not grown to be large, the equations of motion (\ref{hartphieq}), 
(\ref{hartukeq}), and (\ref{Nphieq}) are scale invariant 
with the scaling $\eta \to \mu \eta$, $H \to \mu H$, $t \to t/\mu$, and 
$k \to \mu k$, where $\mu$ is an arbitrary scale.

For completeness, we show the case of the evolution with initial values of the 
Hubble constant given by $H(t_0) = 5$ and $H(t_0) = 2$ respectively in figures
7-8 using radiation dominated expansion. 
Here, we have used the large $N$ approximation and have made an appropriate
increase in the initial value of the zero mode such that the fluctuations (figs.
7b and 8b) grow significantly [we have chosen $\eta(t_0)=40$ in fig. 7 and
$\eta(t_0)=16$ in fig. 8].  While the dynamics looks much like that of 
figs. 4-6 above, we point out that the particle distribution (figs. 7d and 8d)
is extended to higher values with the result being that the equation of state 
(figs. 7e and 8e) is weighted more toward that of radiation.

{\bf Case 3: $m^2>0$, $\eta(t_0)\gg 1$}.  The final case we examine is that of
a simple chaotic scenario with a positive mass term in the Lagrangian.  Again,
preheating can begin only after the inflationary phase of exponential expansion; 
this allows us to take a zero temperature initial state for 
the FRW stage.

Figure 9 shows this situation in the large $N$ approximation for a matter
dominated cosmology.  The zero mode, $\eta(t)$, oscillates in time while
decaying in amplitude from its initial value of $\eta(t_0\! =\! 0)=5$,
$\dot{\eta}(t_0\! =\! 0)=0$ (fig. 9a), while the quantum fluctuation,
$g\Sigma$, grows rapidly for early times due to parametric resonance
(figs. 9b).  We choose here an initial condition on the zero mode which
differs from that of figs 4-5 above since there is no significant growth
of quantum fluctuations for smaller initial values.  From figure 9d, we see
that there exists a single unstable band at values of roughly $k=1$ to
$k=3$, although careful examination reveals that the unstable band 
extends all the way to $k=0$.  The equation of state is depicted by the 
quantity $p(t)/\varepsilon(t)$ in figure 9e. As expected in this massive 
theory, the equation of state is matter dominated.

The final case is the Hartree approximation, shown in figure 10.  Here,
parametric amplification is entirely inefficient when expansion of the universe
is included and we require an initial condition on the zero mode of
$\eta(t_0\! =\! 0)=12\sqrt{3}$ to provide even meager growth of quantum 
fluctuations.  We have used a matter dominated gravitational background with 
$H(t_0)=0.1$.  We see that while the zero mode oscillates (fig. 10a), 
there is little growth in quantum fluctuations (fig. 10b) and few particles
produced (fig. 10c).  Examining the particle distribution (fig. 10d), 
it is found that the bulk of these particles is produced within a single  
resonance band extending from $k \simeq 15$ to $k \simeq 16$.  
This resonance develops at early time during the large amplitude oscillation 
of the zero mode.  These results are explained by a simple resonance band
analysis described below.

At first glace, it is not entirely clear why there are so many more particles
produced in the large $N$ case of figure 9 than in the Hartree case of figure 10.
Since in the present case the Hubble time is long compared to the oscillation
timescale of the zero mode, $H \ll 1$, we would expect a forbidden band for
early times at the location given approximately by the Minkowski results
provided in Ref. \cite{erice}.  In fact, we find this to be the case.  
However, we know from previous studies that in Minkowski space a similar number
of particles is produced in both the Hartree and large $N$ case \cite{mink,erice}.

The solution to this problem is inherent in the band structure of the two cases
when combined with an understanding of the dynamics in an expanding spacetime.
First, we note that, for early times when $g\Sigma \ll 1$, the zero mode 
is well fit by the function $\eta(t)=\eta_0 f(t)/a(t)$ where f(t) is an 
oscillatory function taking on values from $-1$ to $1$.  This is clearly seen
from the envelope function $\eta_0/a(t)$ shown in fig. 10a (recall that 
$g\Sigma \ll 1$ during the entire evolution in this case).  Second, the 
momentum that appears in the equations for the modes (\ref{hartukeq}) is the
{\em physical} momentum $k/a(t)$.  We therefore write the approximate 
expressions for the locations of the forbidden bands in FRW by using the
Minkowski results of \cite{erice} with the substitutions 
$\eta_0^2 \to \gamma\eta_0^2/a^2(t)$ (where the factor of $\gamma$ accounts
for the difference in the definition of the non-linear coupling between
this study and \cite{erice}) and $q^2 \to k^2/a^2(t)$.  

Making these substitutions, we find for the location in comoving momentum $k$ 
of the forbidden band in the large $N$ (fig. 9) and Hartree (fig. 10) cases:
\begin{eqnarray}
0 \leq & k^2 & \leq \frac{\eta_0^2}{2}, \; \; (large N) \\
\frac{\eta_0^2}{2}+3a^2(t) \leq & k^2 & \leq
a^2(t)\left(\sqrt{\frac{\eta_0^4}{3a^4(t)}+\frac{2\eta_0^2}{a^2(t)}+4}+1\right). 
\; \; (Hartree) \label{hartband}
\end{eqnarray}
The important feature to notice is that while the location of the unstable
band (to a first approximation) in the case of the continuous $O(N)$ theory
is the same as in Minkowski and does not change in time, the location of 
the band is time dependent in the discrete theory described by the 
non-perturbative Hartree approximation.

While $\eta_0/a(t) \gg 1$, the Hartree relation reduces to
\begin{equation}
\frac{\eta_0^2}{2} \leq k^2 \leq \frac{\eta_0^2}{\sqrt{3}}.
\end{equation}
This is the same as the Minkowski result for large amplitude, and one finds
that this expression accurately predicts the location of the resonance band
of fig. 10d.  However, with time the band shifts its location toward higher
values of comoving momentum as given by (\ref{hartband}), cutting off particle
production in that initial band.  There is continuing particle production 
for higher modes, but since the Floquet index is decreased due to the reduced
amplitude of the zero mode, since there is no enhancement of production
of particles in these modes (as these modes begin with at most of order $1$
particles), and because the band continues to shift to higher momenta while 
becoming smaller in width, this particle production never becomes significant.

As in the symmetry broken case of figs. 4-6, the equations of motion for
large amplitude and relatively early times are approximately scale invariant.
In figures 11-12 we show the case of the large $N$ evolution in a radiation
dominated universe with initial Hubble constants of $H(t_0)=5$ and $H(t_0)=2$ 
respectively with appropriately scaled initial values of the zero mode of
$\eta(t_0)=40$ and $\eta(t_0)=16$.  Again, the qualitative dynamics remains 
largely unchanged from the case of a smaller Hubble constant.

\subsection{Late Time Behavior}
We see clearly from the numerical evolution that in the case of a symmetry
broken potential, the late time large $N$ solutions obey the sum rule
\begin{equation}
-|m_r^2|+\frac{\lambda_r}{2}\phi^2(t)+
\frac{\lambda_r}{2}\langle\psi^2(t)\rangle_r=0.
\label{sumrule}
\end{equation}
This sum rule is a consequence of the late time Ward identities which enforce
Goldstone's Theorem.  Because of this sum rule, we can write down the
analytical expressions for the late time behavior of the fluctuations and the
zero mode.  Using (\ref{sumrule}), the mode equation (\ref{hartukeq}) becomes
\begin{equation}
\left[\frac{d^2}{dt^2}+3\frac{\dot{a}(t)}{a(t)}\frac{d}{dt}+\frac{k^2}{a^2(t)}
\right]U_k(t) = 0.
\label{lateuk}
\end{equation}
This equation can be solved exactly if we assume a power law dependence for the
scale factor $a(t) = (t/t_0)^n$.  The solution is
\begin{equation}
U_k(t) = c_k \; t^{(1-3n)/2} \; J_{\frac{1-3n}{2-2n}}\left(\frac{k t_0^n
	t^{1-n}}{n-1}\right) + 
	d_k \; t^{(1-3n)/2} \; Y_{\frac{1-3n}{2-2n}}\left(\frac{k t_0^n
	t^{1-n}}{n-1}\right), 
\label{latesoln}
\end{equation}
where $J_{\nu}$ and $Y_{\nu}$ are Bessel and Neumann functions respectively, 
and the constants $c_k$ and $d_k$ 
carry dependence on the initial conditions and on the dynamics up to the point
at which the sum rule is satisfied.  

These functions have several important properties.  In particular, in radiation
or matter dominated universes, $n<1$, and for values of wavenumber satisfying
$k \gg t^{-(1-n)}/t_0^n$, the mode functions decay in time as 
$1/a(t) \sim t^{-n}$.  Since
the sum rule applies for late times, $t-t_0 \gg 1$ in dimensionless units, we see
that all values of $k$ except a very small band about $k=0$ redshift as
$1/a(t)$.  The $k=0$ mode, however, remains constant in time, explaining the
support evidenced in the numerical results for values of small $k$ (see
figs. 1,3).  These results mean that the quantum fluctuation has a late time
dependence of $\langle \psi^2(t)\rangle_r \sim 1/a^2(t)$.  The late time
dependence of the zero mode is given by this expression combined with the sum
rule (\ref{sumrule}).  These results are accurately reproduced by our numerical
analysis.  Note that qualitatively this late time dependence is independent of
the choice of initial conditions for the zero mode, except that there is no
growth of modes near $k=0$ in the case in which particles are produced via
parametric amplification (figs. 4,5).

For the radiation $ n= \frac12 $ and matter dominated $ n = \frac23 $
universes, eq.(\ref{latesoln}) reduces to elementary functions:
\begin{eqnarray}
a(t) \; U_k(t) &=&  c_k \; e^{2ik t_0^{1/2} t^{1/2} } + 
d_k\; e^{-2ik t_0^{1/2} t^{1/2} }  
\;  \; \mbox{(RD) } \; ,\cr \cr
a(t) \; U_k(t) &=&  c_k \; e^{3ik t_0^{2/3} t^{1/3} } \; 
\left[ 1 + {i \over {3k t_0^{2/3}
t^{1/3}}}\right] + d_k\; e^{-3ik t_0^{2/3} t^{1/3}}\left[ 1 - {i \over {3k
t_0^{2/3} t^{1/3}}}\right] \; \; \mbox{ (MD) }.
\end{eqnarray}

It is also of interest to examine the $n>1$ case.  Here, the modes of interest
satisfy the condition $k \ll t^{n-1}/t_0^n$ for late times.  These modes are constant
in time and one sees that the modes are {\it frozen}.  In the case of a de
Sitter universe, we can formally take the limit $n \to \infty$ and we see that
{\it all} modes become frozen at late times.  This case was studied in detail 
in ref. \cite{usDeS}.

\section{Conclusions}

We have shown that there can be significant particle production through quantum
fluctuations after inflation.  However, this production is somewhat sensitive
to the expansion of the universe. From our analysis of the equation of state,
we see that the late time dynamics is given by a matter dominated cosmology.
We have also shown that the quantum fluctuations of the inflaton decay for late
times as $1/a^2(t)$, while in the case of a symmetry broken inflationary model,
the inflaton field moves to the minimum of its tree level potential.  The
exception to this behavior is the case when the inflaton begins exactly at the
unstable extremum of its potential for which the fluctuations grow out to the
minimum of the potential and do not decay.  Initial production of particles due
to parametric amplification is significantly greater in chaotic scenarios with
symmetry broken potentials than in the corresponding theories with positive
mass terms in the Lagrangian, given similar initial conditions on the zero mode
of the inflaton.

Since there are a number of articles in the literature treating the problem of
preheating, it is useful to review the unique features of the present work.
First, we have treated the problem {\em dynamically}, without using the
effective potential (an equilibrium construct) to determine the evolution.
Second, we have provided consistent non-perturbative calculations of the
evolution to bring out some of the most relevant aspects of the late time
behavior.  In particular, we found that the quantum backreaction naturally
inhibits catastrophic growth of fluctuations and provides a smooth transition
to the late time regime in which the quantum fluctuations decay as the zero
mode approaches its asymptotic state.  Third, the dynamics studied obeys the
constraint of covariant conservation of the energy momentum tensor.

The next stage of this analysis is to allow dynamical evolution of the scale
factor, to follow the evolution of the inflaton through the de Sitter stage and
into the stage of particle production.  As the expansion rates for which there
is significant particle production is somewhat restrictive, it is
yet to be seen whether in a theory with fully dynamical gravitational expansion
such particle production will be a significant factor.  This analysis is
currently underway \cite{grav}.

\acknowledgements 

D.B. thanks the NSF for support under grants PHY-9302534 and
INT-9216755. R.H. and D.C. were supported by DOE grant
DE-FG02-91-ER40682. A.S. and M.S. were supported by the NSF under grant
PHY-91-16964.

\appendix
\section{Conformal time analysis and initial Conditions}

The issue of renormalization and initial conditions is best understood in
conformal time which is a natural framework for adiabatic renormalization and
regularization.

Quantization in conformal time proceeds by writing the metric element as
\begin{equation}
ds^2= C^2(\tau)(d\tau^2 - d\vec{x}^2). 
\end{equation}
Under a conformal rescaling of the field
\begin{equation}
\Phi(\vec x, t) = \chi(\vec x, \tau)/ C(\tau),
\end{equation}
the action for a scalar field (with the obvious generalization to N
components) becomes, after an integration by parts and dropping a
 surface term
\begin{equation}
S= \int d^3x d\tau \left\{\frac12 (\chi')^2-\frac12 (\vec{\nabla}\chi)^2-
{\cal{V}}(\chi)\right\},
\end{equation}
with
\begin{equation}
{\cal{V}}(\chi) = C^4(\tau)V(\Phi/C(\tau))-C^2(\tau)\frac{{\cal{R}}}{6}\chi^2,
\end{equation}
where ${\cal{R}}= 6C''(\tau)/C^3(\tau)$ is the Ricci scalar,
and primes stand for derivatives with respect to conformal time $\tau$.

The conformal time Hamiltonian operator, which is the generator of translations
in $\tau$, is given by
\begin{equation}
H_{\tau}= \int d^3x \left\{ \frac{1}{2}\Pi^2_{\chi}+\frac{1}{2}
(\vec{\nabla}\chi)^2+{\cal{V}}(\chi) \right \}, \label{confham}
\end{equation}
with $\Pi_{\chi}$ being the canonical momentum conjugate to $\chi$,
$\Pi_{\chi} = \chi'$. 
Separating the zero mode of the field $\chi$ 
\begin{equation}
\chi(\vec x, \tau) = \chi_0(\tau) + \bar{\chi}(\vec x,\tau),
\end{equation}
and performing the large $N$ or Hartree factorization on the
fluctuations we find 
that the Hamiltonian becomes linear plus quadratic in the fluctuations, and
similar to a Minkowski space-time Hamiltonian with a $\tau$ dependent mass term
given by
\begin{equation}
{\cal{M}}^2(\tau) = C^2(\tau) \left[m^2+(\xi-\frac{1}{6}){\cal{R}}
+ \frac{\lambda}{2}\chi_0^2(\tau) + \frac{\lambda}{2}
\langle \bar{\chi}^2 \rangle\right]. \label{masseff}
\end{equation}

We can now follow the steps and use the results of reference\cite{frw} for the
conformal time evolution of the density matrix by setting $a(t)=1$ in the
proper equations of that reference and replacing the frequencies by
\begin{equation}
\omega^2_k(\tau) = \vec{k}^2 + {\cal{M}}^2(\tau), \label{freqs}
\end{equation}
and the expectation value in (\ref{masseff}) is obtained in this
$\tau$ evolved density matrix.
The time evolution of the kernels in the density matrix (see\cite{frw})
is determined by the mode functions that obey
\begin{equation}
\left[ \frac{d^2}{d\tau^2}+k^2+{\cal{M}}^2(\tau)\right] f_k(\tau)=0.
\label{fmodeqn}
\end{equation}
The Wronskian of these mode functions
\begin{equation}
{\cal{W}}(f,f^*)= f'_k f^*_k-f_k f^{*'}_k
\end{equation} 
is a constant. It is natural to impose initial conditions such that at the
initial $\tau$ the density matrix describes a situation of local thermodynamic
equilibrium and therefore commutes with the conformal time Hamiltonian at the
initial time. This implies that the initial conditions of the mode functions
$f_k(\tau)$ should be chosen to be (see\cite{frw})
\begin{equation}
f_k(\tau_o)= \frac{1}{\sqrt{\omega_k(\tau_o)}} \; \; ; \; 
f'_k(\tau_o)= -i\sqrt{\omega_k(\tau_o)} f_k(\tau_o). \label{inicond}
\end{equation}
These initial conditions correspond to the choice of mode functions which
coincide with the first order adiabatic modes and those of the Bunch-Davies
vacuum for large momentum\cite{birrell}.  To see this clearly, consider the WKB
solutions of the mode equation (\ref{fmodeqn}) of the form
\begin{equation}
D_k(\tau) = e^{\int^{\tau}_{\tau_o} R_k(\tau')d\tau'},
\end{equation} 
with the function $R_k(\tau)$ obeying the Riccati equation
\begin{equation}
R'_k + R^2_k+ k^2+{\cal{M}}^2(\tau)=0.
\end{equation} 
This equation possesses the solution
\begin{equation}
R_k(\tau) = -ik+R_{0,k}(\tau)-i\frac{R_{1,k}(\tau)}{k}+
\frac{R_{2,k}(\tau)}{k^2}-i\frac{R_{3,k}(\tau)}{k^3}+
\frac{R_{4,k}(\tau)}{k^4}+ \cdots 
\end{equation}
and its complex conjugate. We find for the  coefficients:
\begin{eqnarray}
&& R_{0,k} = 0 \; \; ; \; \; R_{1,k} = \frac{1}{2} {\cal{M}}^2(\tau) \; \; ; \;
\; R_{2,k} = -\frac{1}{2} R'_{1,k} \nonumber \\ &&R_{3,k} = \frac{1}{2}\left(
R'_{2,k}-R^2_{1,k} \right) \; \; ; \; \; R_{4,k} = -\frac{1}{2}\left(
R'_{3,k}+2R_{1,k}R_{2,k} \right).
\end{eqnarray}
The solutions $f_k(\tau)$ obeying the boundary conditions (\ref{inicond}) are
obtained as linear combinations of this WKB solution and its complex conjugate
\begin{equation}
f_k(\tau) = \frac{1}{2\sqrt{\omega_k(\tau_o)}}\left[
(1+\gamma)D_k(\tau)+(1-\gamma)D^*_k(\tau) \right],
\end{equation}
where the coefficient $\gamma$ is obtained from the initial conditions. It is
straightforward to find that the real and imaginary parts are given by
\begin{equation}
\gamma_R= 1+ {\cal{O}}(1/k^4) \; \; ; \; \; \gamma_I= {\cal{O}}(1/k^3).
\end{equation}
Therefore the large-$k$ mode functions satisfy the adiabatic vacuum initial
conditions\cite{birrell}. This, in fact, is the rationale for the choice of the
initial conditions (\ref{inicond}).

Following the analysis presented in \cite{frw} we find, in conformal time that
\begin{equation}
\langle \bar{\chi}^2(\vec x,\tau) \rangle = \int \frac{d^3k}{2(2\pi)^3}
\; |f_k(\tau)|^2.
\end{equation}
The Heisenberg field operators $\bar{\chi}(\vec x, \tau)$  and 
their canonical momenta $\Pi_{\chi}(\vec x, \tau)$ can be expanded as:
\begin{eqnarray}
&& \bar{\chi}(\vec x, \tau) = \int {{d^3k}\over {\sqrt2 \, (2\pi)^{3/2}}} 
\left[ a_k f_k(\tau)+ a^{\dagger}_{-k}f^*_k(\tau) \right]
 e^{i \vec k \cdot \vec x}, \label{heisop}\\
&& \Pi_{\chi}(\vec x, \tau) = 
\int {{d^3k}\over {\sqrt2 \, (2\pi)^{3/2}}} 
\left[ a_k f'_k(\tau)+ a^{\dagger}_{-k}f^{*'}_k(\tau) \right]
 e^{i \vec k \cdot \vec x}, \label{canheisop}
\end{eqnarray}
with the time independent creation and
annihilation operators $ a_k$ and $ a^{\dagger}_k $ 
obeying canonical commutation relations. Since the fluctuation fields
in comoving and conformal time are related by a conformal rescaling
\begin{equation}
\psi(\vec x, t) = \frac{\chi(\vec x, \tau)}{C(\tau)},
\end{equation}
it is straightforward to see that the mode functions in comoving time are
related to those in conformal time simply as
\begin{equation}
U_k(t) = \frac{f_k(\tau)}{C(\tau)}.
\end{equation}
Therefore the initial conditions (\ref{inicond}) on the conformal time mode
functions imply the initial conditions for the mode functions in comoving time
are given by
\begin{equation}
U_k(t_0) = \frac{1}{\omega_k(\tau_0)} \; \; ; \; \dot{U}_k(t_0)=
\left[-i\omega_k(\tau_0)-H(t_0)\right] U_k(t_0), \label{inicondcomo}
\end{equation}
where we have chosen the normalization of the scale factor such that
$a(t_0)= C(\tau_0)=1$. 
 
For renormalization purposes we need the large-$k$ behavior of $|U_k(t)|^2 \; ,
|\dot{U}_k(t)|^2$, which are determined by the large-$k$ behavior of the
conformal time mode functions and its derivative. These are given by
\begin{equation}
|f_k(\tau)|^2 = \frac{1}{k}\left[ 1-\frac{R_{1,k}(\tau)}{k^2}+
\frac{1}{k^4}\left(\frac{R''_{1,k}(\tau)}{4}+\frac{3}{2}R^2_{1,k}(\tau)
\right ) 
+\cdots \right], \label{largekf}
\end{equation}

\begin{equation}
|f'_k(\tau)|^2 = {k}\left[ 1+\frac{R_{1,k}(\tau)}{k^2}+
\frac{1}{k^4}\left(-\frac{R''_{1,k}(\tau)}{4}+\frac{3}{2}R^2_{1,k}(\tau) \right
) +\cdots \right]. \label{largekfprime}
\end{equation}

We note that the large $k$ behavior of the mode functions to the order needed to
renormalize the quadratic and logarithmic divergences is insensitive to the
initial conditions. This situation must be contrasted with the case in which
the initial conditions in comoving time are imposed as described
in\cite{frw,usDeS}. Thus the merit in considering the initial conditions in
conformal time described in this article.

The correspondence with the comoving time mode functions is given by:
\begin{eqnarray}
|U_k(t)|^2 & = & \frac{|f_k(\tau)|^2}{C^2(\tau)} \nonumber \\
|\dot{U}_k(t)|^2 & = & \frac{1}{C^2(\tau)}
\left[ \frac{|f'_k(\tau)|^2}{C^2(\tau)}+ \left(H^2-\frac{H}{C(\tau)}
\frac{d}{d\tau}\right)|f_k(\tau)|^2 \right]
\end{eqnarray}
These are asymptotic forms used in the renormalization program in section III.

There is an important physical consequence of this choice of initial
conditions, which is revealed by analyzing the evolution of the density matrix.

In the large $N$ or Hartree (also to one-loop) approximation, the
density matrix 
is Gaussian, and defined by a normalization factor, a complex covariance that
determines the diagonal matrix elements and a real covariance that determines
the mixing in the Schr\"odinger representation as discussed in
reference\cite{frw} (and references therein).

In conformal time quantization and in the Schr\"odinger representation in which
the field $\chi$ is diagonal the conformal time evolution of the density matrix
is via the conformal time Hamiltonian (\ref{confham}). The evolution equations
for the covariances is obtained from those given in reference\cite{frw} by
setting $a(t) = 1$ and using the frequencies $\omega^2_k(\tau)=
k^2+{\cal{M}}^2(\tau)$. In particular, by setting the covariance of the
diagonal elements (given by equation (2.20) in\cite{frw}; see also equation
(2.44) of\cite{frw}),
\begin{equation}
{\cal{A}}_k(\tau) = -i \frac{f^{'*}_k(\tau)}{f^*_k(\tau)},
\end{equation}
we find that with the initial conditions (\ref{inicond}), the
conformal time density matrix is that of local equilibrium at $\tau_0$
in the sense that it commutes with the conformal time Hamiltonian. 
However, it is straightforward to see, that the comoving time density
matrix {\em does not} commute with the {\it comoving time} Hamiltonian at
the initial time $t_0$.  

An important corollary of this analysis and comparison with other initial
conditions used in comoving time is that assuming initial conditions of local
equilibrium in comoving time leads to divergences that depend on the initial
condition as discussed at length in\cite{frw}.  This dependence of the
renormalization counterterms on the initial condition was also realized by
Leutwyler and Mallik\cite{leutwyler} within the context of the CTP formulation.
Imposing the initial conditions corresponding to local thermal equilibrium in
{\em conformal} time, we see that: i) the renormalization counterterms do not
depend on the initial conditions and ii) the mode functions are identified with
those corresponding to the adiabatic vacuum for large momenta.

Thus this analysis justifies the use of the initial conditions on the {\em
comoving} mode functions (\ref{inicondcomo}). Furthermore, being that the
comoving time density matrix does not describe at any time a condition of local
thermodynamic equilibrium, the ``temperature'' that enters in the mixing
covariance in the density matrix is understood as a parameter describing a
mixed state with the notion of temperature at local thermodynamic equilibrium
in conformal time quantization.

For our main analysis we choose this ``temperature'' to be zero so that the
resulting density matrix describes a pure state, which for the large
momentum modes coincides with the conformal adiabatic vacuum.

{\bf Particle Number:}

We write the Fourier components of the field $\chi$ and its canonical
momentum $\Pi_{\chi}$ given by (\ref{heisop}) -(\ref{canheisop}) as:
\begin{eqnarray}
&& \bar{\chi}_k(\tau) = \frac{1}{\sqrt{2}}\left[
a_k f_k(\tau)+ a^{\dagger}_{-k}f^*_k(\tau) \right],
  \label{heisopfu}\\
&& \Pi_{\chi,k}(\tau) = \frac{1}{2}\left[
a_k f'_k(\tau)+ a^{\dagger}_{-k}f^{*'}_k(\tau) \right].
  \label{canheisopfu}
\end{eqnarray}
These (conformal time) Heisenberg operators can be written equivalently
in terms of the $\tau$ dependent creation and annihilation operators

\begin{eqnarray}
&& \bar{\chi}_k(\tau) = \frac{1}{\sqrt{2 \omega_k(\tau_0)}}\left[
\tilde{a}_k(\tau) e^{-i\omega_k(\tau_0)\tau}+ 
\tilde{a}^{\dagger}_k(\tau) e^{i\omega_k(\tau_0)\tau}
 \right],
  \label{newheis}\\
&& \Pi_{\chi,k}(\tau) = -i\sqrt{\frac{\omega_k(\tau_0)}{2}}\left[
\tilde{a}_k(\tau) e^{-i\omega_k(\tau_0)\tau}-
\tilde{a}^{\dagger}_k(\tau) e^{i\omega_k(\tau_0)\tau}
 \right].
  \label{newcanheisopfu}
\end{eqnarray}

The operators $\tilde{a}_k(\tau) \; ; a_k(\tau) $ are related by a 
Bogoliubov transformation. The number of particles referred to the
initial Fock vacuum of the modes $f_k$, is given by
\begin{equation}
N_k(\tau) = \langle \tilde{a}^{\dagger}_k(\tau) \tilde{a}_k(\tau) \rangle 
= \frac{1}{4} \left|\frac{f_k(\tau)}{f_k(0)}\right|^2 \left [
1+ \frac{1}{\omega^2_k(\tau_0)} \left| \frac{f'_k(\tau)}{f_k(\tau)} 
\right|^2 \right] - \frac{1}{2} \; , \label{partnumb}
\end{equation}
 or alternatively, in terms of the comoving mode functions $U_k(t) =
f_k(\tau)/C(\tau)$ we find

\begin{equation}
N_k(t) = \frac{a^2(t)}{4} \left| \frac{U_k(t)}{U_k(0)}\right|^2
\left[ 1+ \frac{1}{\omega^2_k(0)} \left| \frac{ \dot{U}_k(t) + H U_k(t)}{U_k(t)}
\right|^2 \right] - \frac{1}{2} \; . \label{partnumbcomo}
\end{equation}

Using the large $k$-expansion of the conformal mode functions given by
eqns. (\ref{largekf}) -(\ref{largekfprime}) we find the large-$k$ behavior
of the particle number to be $N_k \buildrel {k\to \infty} \over =
{\cal{O}}(1/k^4) $, and the 
total number of particles (with reference to the initial state at $\tau_0$) 
is therefore finite.

\newpage

\begin{figure}
\epsfig{file=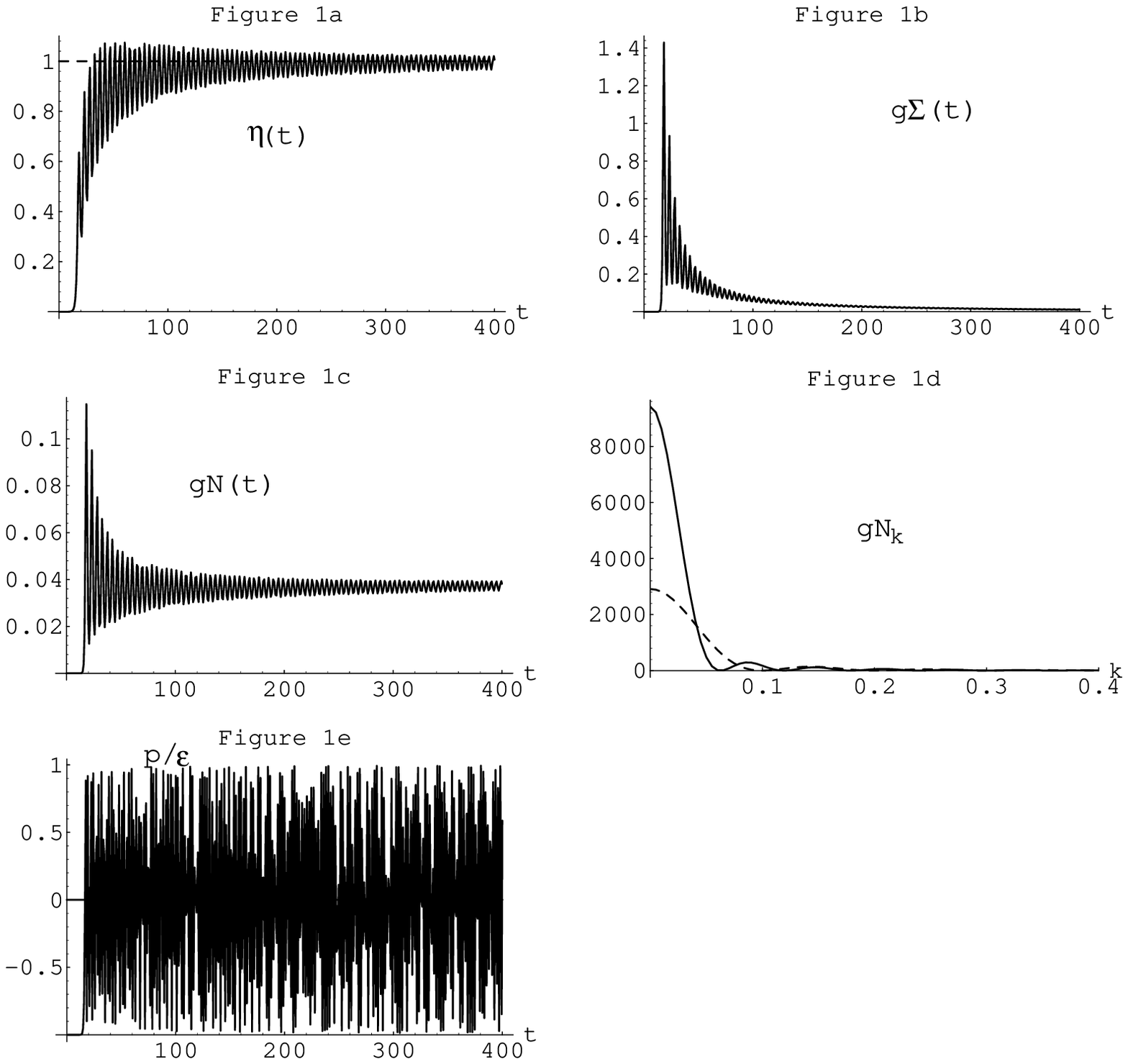}
\caption{Figure 1: Symmetry broken, slow roll, large $N$, matter dominated
evolution of (a) the zero mode $\eta(t)$ vs. $t$, (b) the quantum fluctuation
operator $g\Sigma(t)$ vs. $t$, (c) the number of particles $gN(t)$ vs. $t$,
(d) the particle distribution $gN_k(t)$ vs. $k$ at $t=149.1$ (dashed line)
and $t=398.2$ (solid line),  and (e) the ratio of the pressure and energy density
$p(t)/\varepsilon(t)$ vs. $t$ for the parameter values $m^2=-1$, $\eta(t_0) =
10^{-7}$, $\dot{\eta}(t_0)=0$, $g = 10^{-12}$, $H(t_0) = 0.1$. \label{fig1}}
\end{figure}

\begin{figure}
\epsfig{file=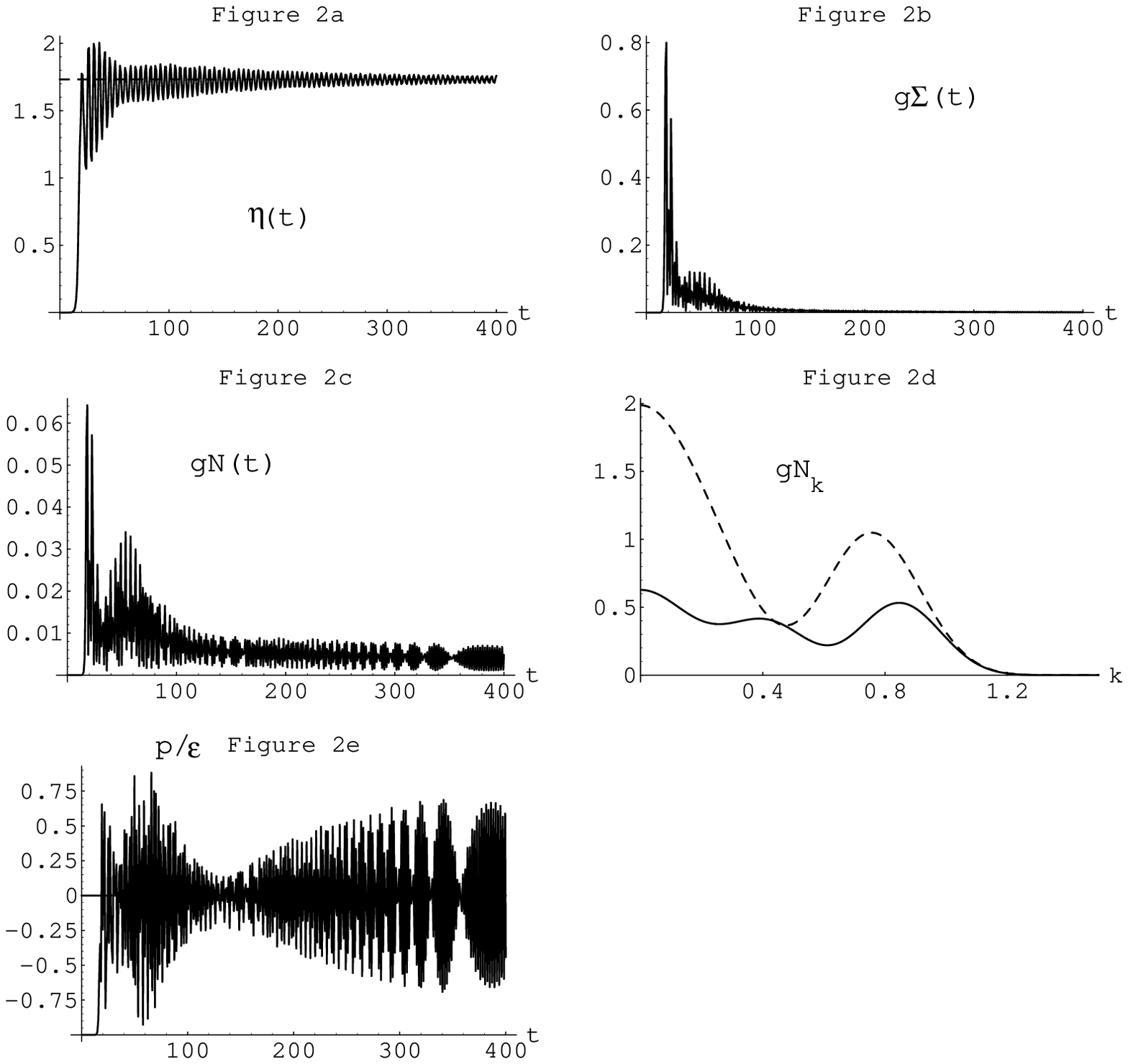}
\caption{Figure 2: Symmetry broken, slow roll, Hartree, matter dominated
evolution of (a) the zero mode $\eta(t)$ vs. $t$, (b) the quantum fluctuation
operator $g\Sigma(t)$ vs. $t$, (c) the number of particles $gN(t)$ vs. $t$,
(d) the particle distribution $gN_k(t)$ vs. $k$ at $t=150.7$ (dashed line)
and $t=396.1$ (solid line),  and (e) the ratio of the pressure and energy density
$p(t)/\varepsilon(t)$ vs. $t$ for the parameter values $m^2=-1$, $\eta(t_0) =
3^{1/2}\cdot 10^{-7}$, $\dot{\eta}(t_0)=0$, $g = 10^{-12}$, $H(t_0) =
0.1$. \label{fig2}}
\end{figure}

\begin{figure}
\epsfig{file=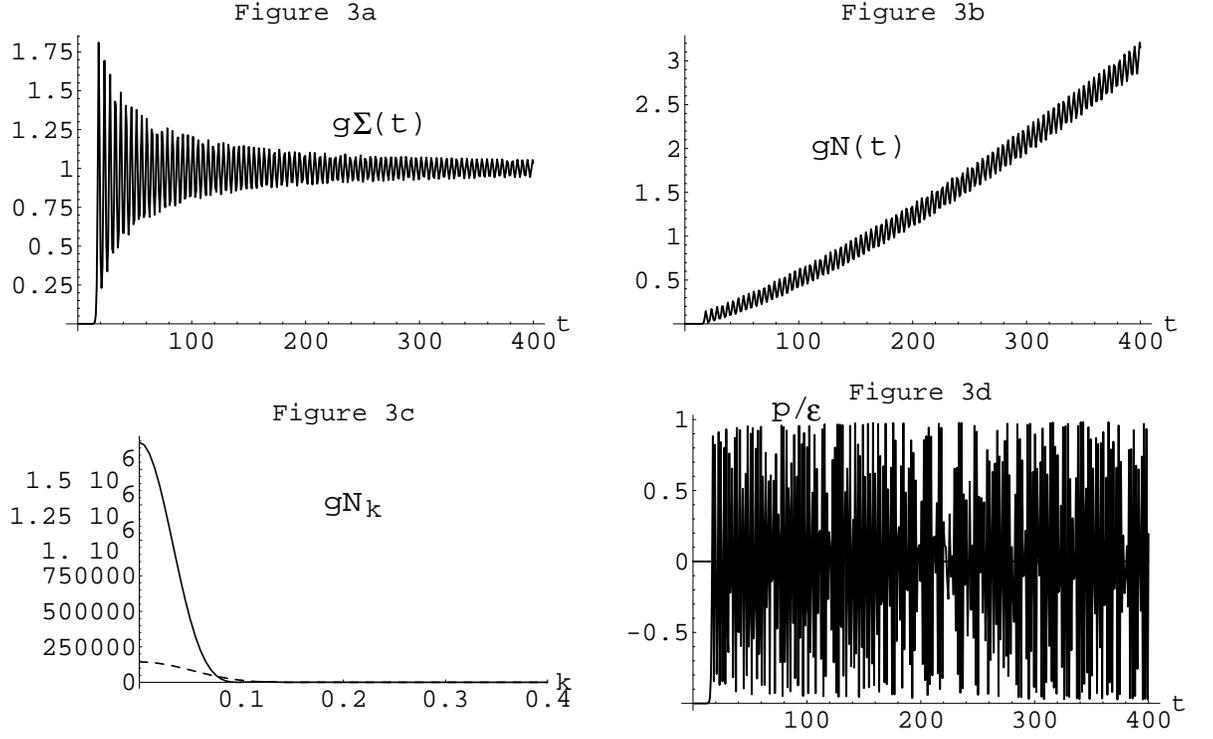}
\caption{Figure 3: Symmetry broken, no roll, matter dominated
evolution of (a) the quantum fluctuation operator $g\Sigma(t)$ vs. $t$, 
(b) the number of particles $gN(t)$ vs. $t$,
(c) the particle distribution $gN_k(t)$ vs. $k$ at $t=150.1$ (dashed line)
and $t=397.1$ (solid line),  and (d)
the ratio of the pressure and energy density $p(t)/\varepsilon(t)$ vs. $t$ for
the parameter values $m^2=-1$, $\eta(t_0) = 0$, $\dot{\eta}(t_0)=0$, $g =
10^{-12}$, $H(t_0) = 0.1$. \label{fig3}}
\end{figure}

\begin{figure}
\epsfig{file=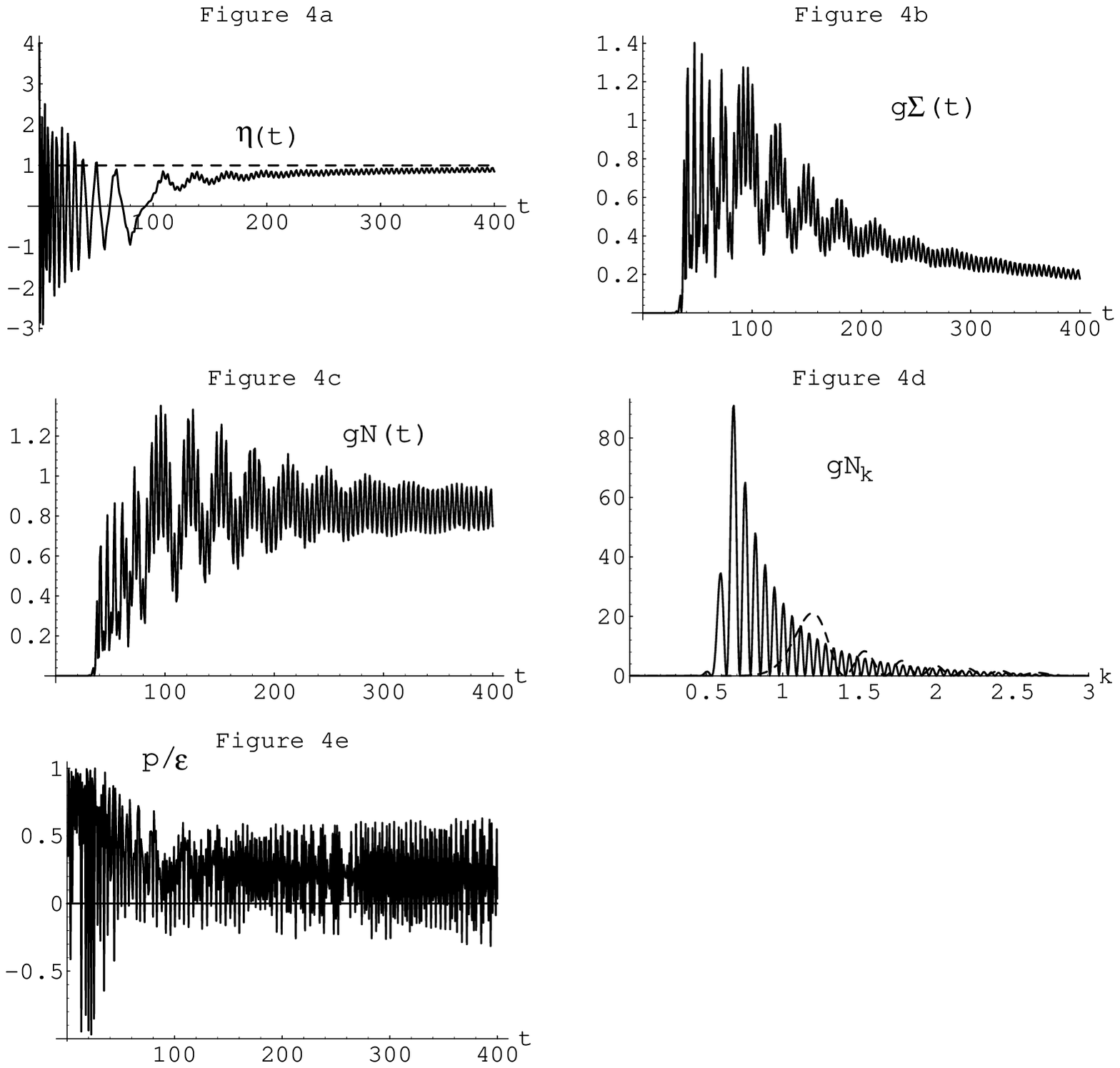}
\caption{Figure 4: Symmetry broken, chaotic, large $N$, radiation dominated
evolution of (a) the zero mode $\eta(t)$ vs. $t$, (b) the quantum fluctuation
operator $g\Sigma(t)$ vs. $t$, (c) the number of particles $gN(t)$ vs. $t$,
(d) the particle distribution $gN_k(t)$ vs. $k$ at $t=76.4$ (dashed line)
and $t=392.8$ (solid line),  and (e) the ratio of the pressure and energy
density $p(t)/\varepsilon(t)$ vs. $t$ for the parameter values $m^2=-1$,
$\eta(t_0) = 4$, $\dot{\eta}(t_0)=0$, $g = 10^{-12}$, $H(t_0) =
0.1$. \label{fig4}}
\end{figure}

\begin{figure}
\epsfig{file=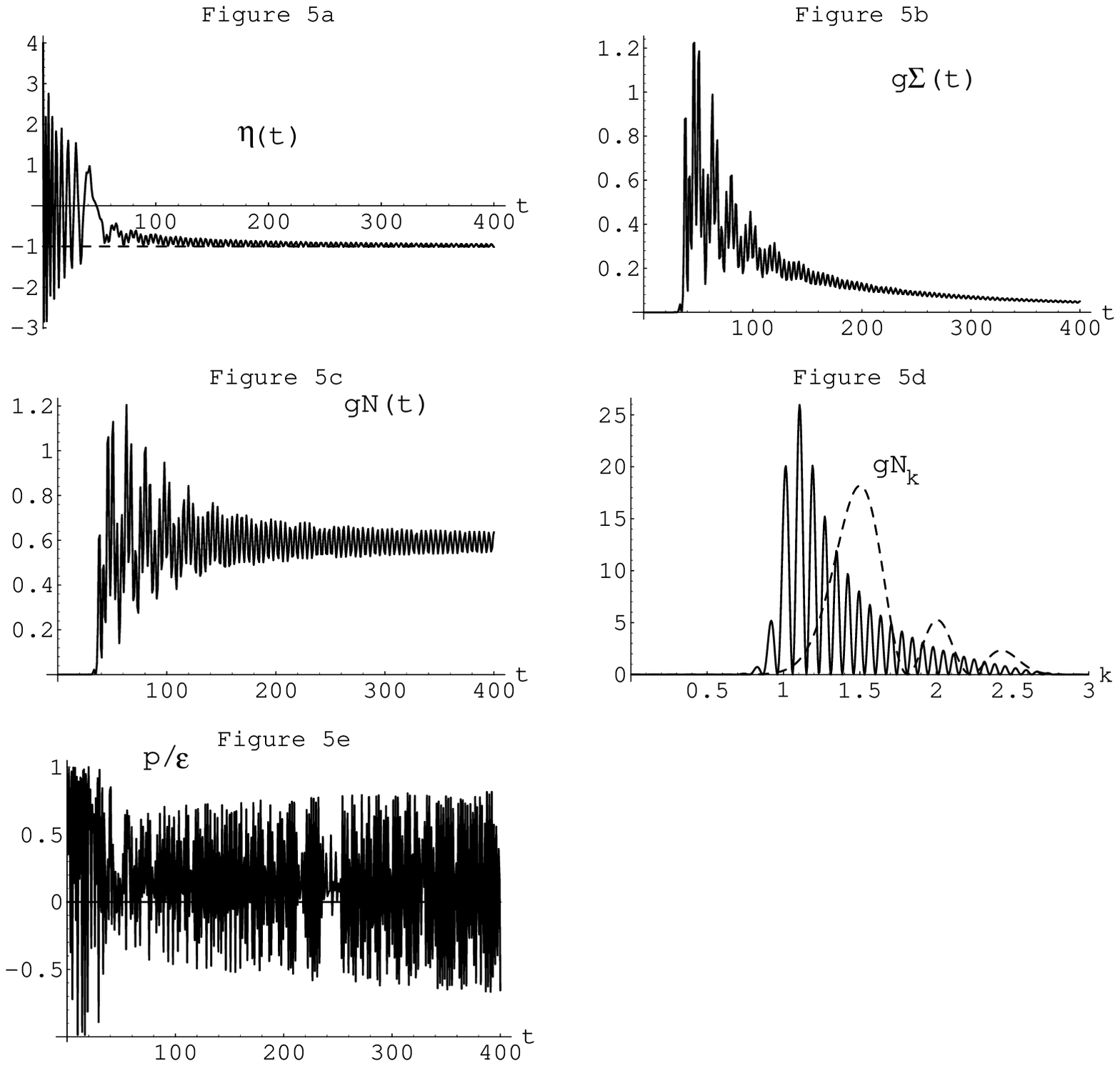}
\caption{Figure 5: Symmetry broken, chaotic, large $N$, matter dominated
evolution of (a) the zero mode $\eta(t)$ vs. $t$, (b) the quantum fluctuation
operator $g\Sigma(t)$ vs. $t$, (c) the number of particles $gN(t)$ vs. $t$,
(d) the particle distribution $gN_k(t)$ vs. $k$ at $t=50.8$ (dashed line)
and $t=399.4$ (solid line),  and (e) the ratio of the pressure and energy density
$p(t)/\varepsilon(t)$ vs. $t$ for the parameter values $m^2=-1$, $\eta(t_0) =
4$, $\dot{\eta}(t_0)=0$, $g = 10^{-12}$, $H(t_0) = 0.1$. \label{fig5}}
\end{figure}

\begin{figure}
\epsfig{file=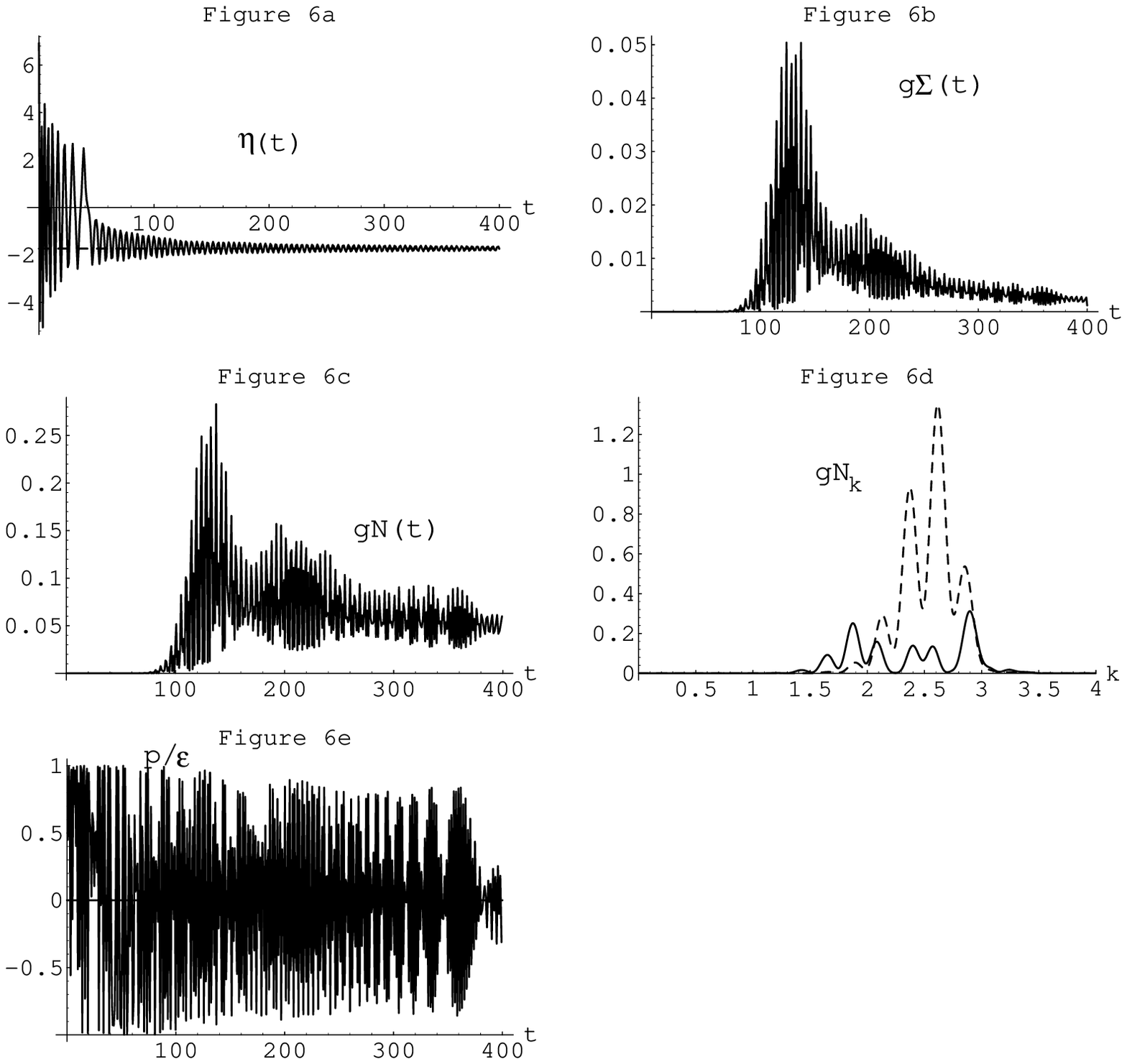}
\caption{Figure 6: Symmetry broken, chaotic, Hartree, matter dominated
evolution of (a) the zero mode $\eta(t)$ vs. $t$, (b) the quantum fluctuation
operator $g\Sigma(t)$ vs. $t$, (c) the number of particles $gN(t)$ vs. $t$,
(d) the particle distribution $gN_k(t)$ vs. $k$ at $t=151.3$ (dashed line)
and $t=397.0$ (solid line),  and (e) the ratio of the pressure and energy density
$p(t)/\varepsilon(t)$ vs. $t$ for the parameter values $m^2=-1$, $\eta(t_0) =
4\cdot 3^{1/2}$, $\dot{\eta}(t_0)=0$, $g = 10^{-12}$, $H(t_0) =
0.1$. \label{fig6}}
\end{figure}

\begin{figure}
\epsfig{file=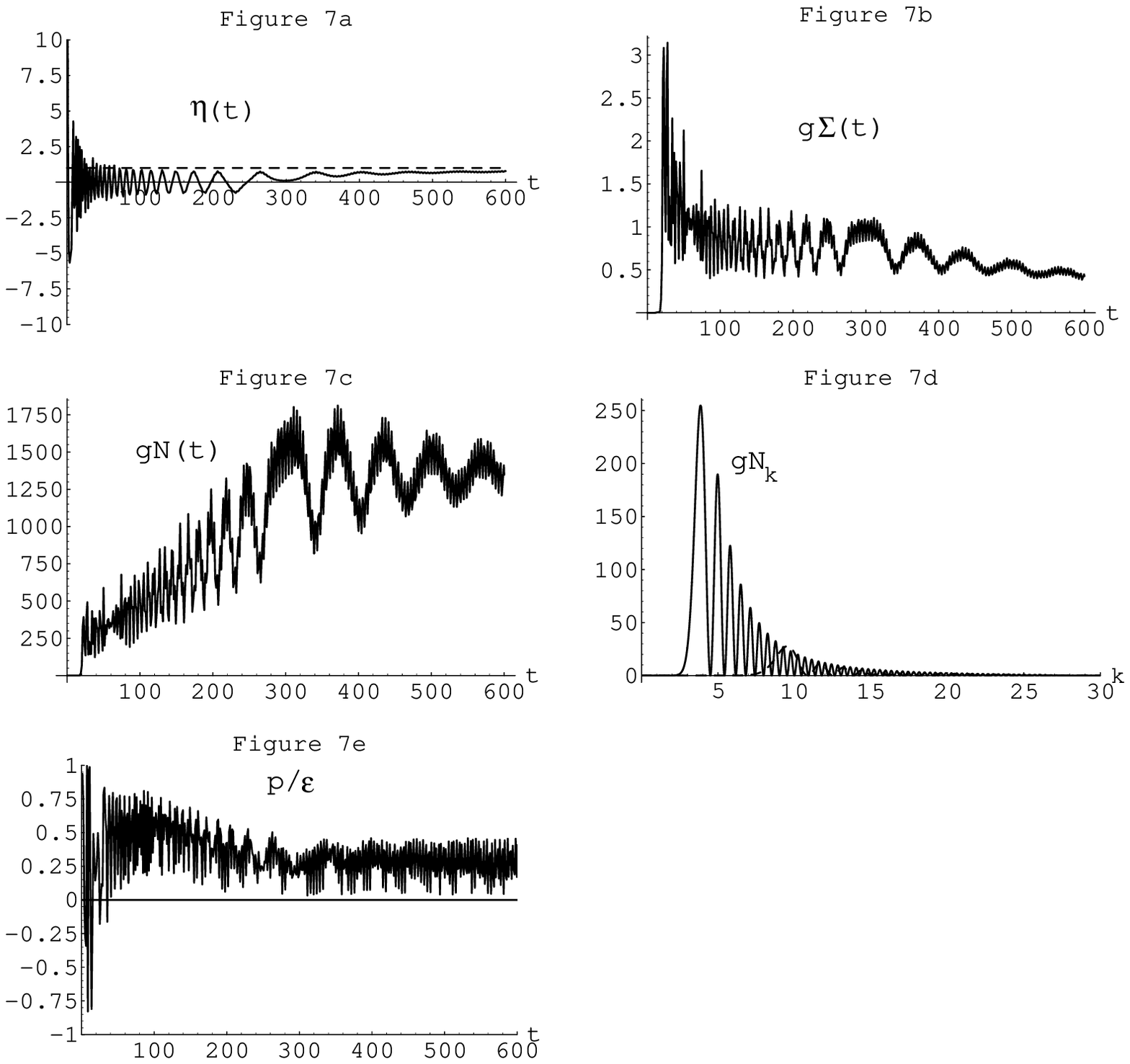}
\caption{Figure 7: Symmetry broken, chaotic, large $N$, radiation dominated
evolution of (a) the zero mode $\eta(t)$ vs. $t$, (b) the quantum fluctuation
operator $g\Sigma(t)$ vs. $t$, (c) the number of particles $gN(t)$ vs. $t$,
(d) the particle distribution $gN_k(t)$ vs. $k$ at $t=118.9$ (dashed line)
and $t=394.7$ (solid line),  and (e) the ratio of the pressure and energy density
$p(t)/\varepsilon(t)$ vs. $t$ for the parameter values $m^2=-1$, $\eta(t_0) =
40$, $\dot{\eta}(t_0)=0$, $g = 10^{-12}$, $H(t_0) =
5.0$. \label{fig7}}
\end{figure}

\begin{figure}
\epsfig{file=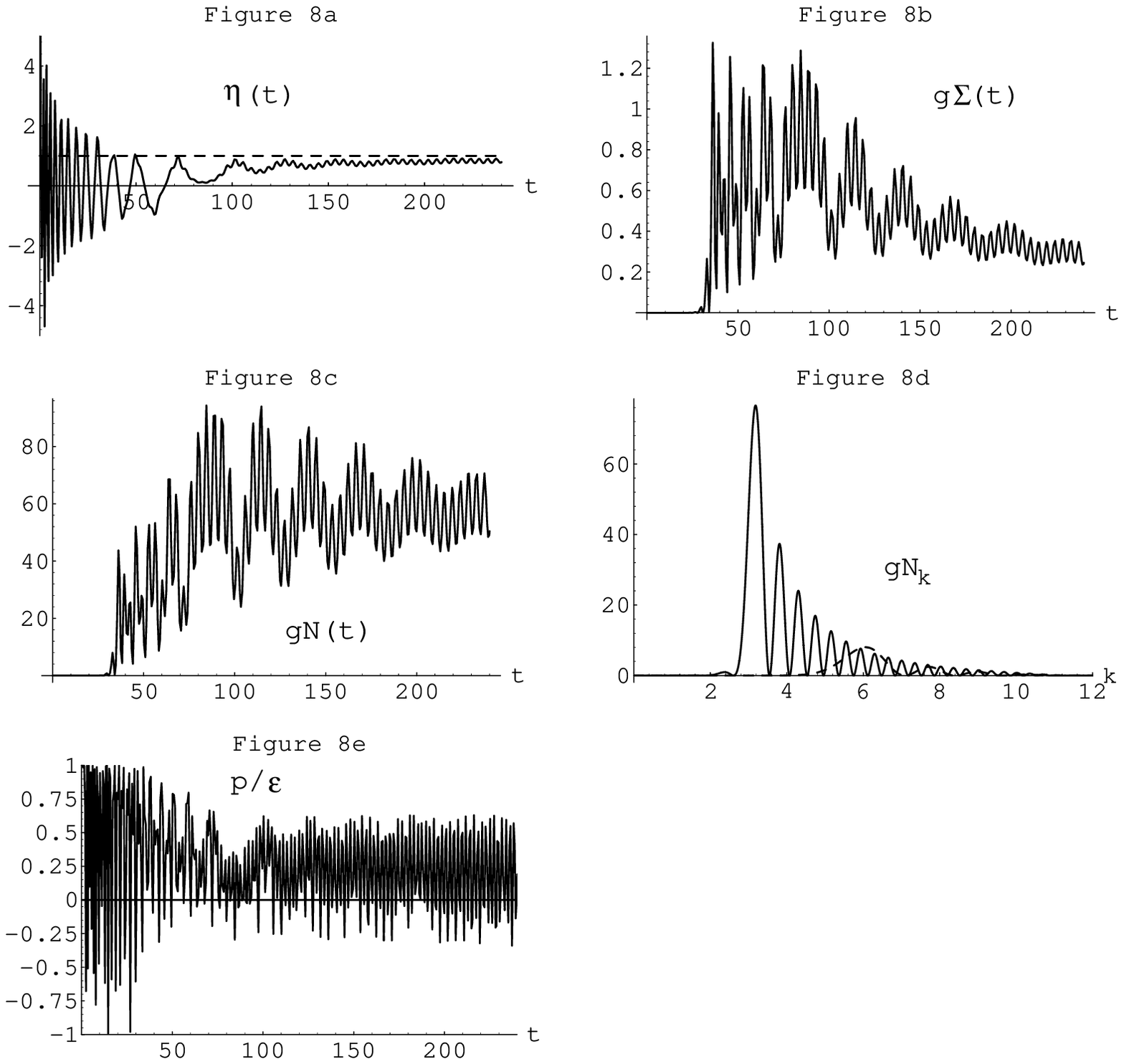}
\caption{Figure 8: Symmetry broken, chaotic, large $N$, radiation dominated
evolution of (a) the zero mode $\eta(t)$ vs. $t$, (b) the quantum fluctuation
operator $g\Sigma(t)$ vs. $t$, (c) the number of particles $gN(t)$ vs. $t$,
(d) the particle distribution $gN_k(t)$ vs. $k$ at $t=55.1$ (dashed line)
and $t=194.2$ (solid line),  and (e) the ratio of the pressure and energy density
$p(t)/\varepsilon(t)$ vs. $t$ for the parameter values $m^2=-1$, $\eta(t_0) =
16$, $\dot{\eta}(t_0)=0$, $g = 10^{-12}$, $H(t_0) =
2.0$. \label{fig8}}
\end{figure}

\begin{figure}
\epsfig{file=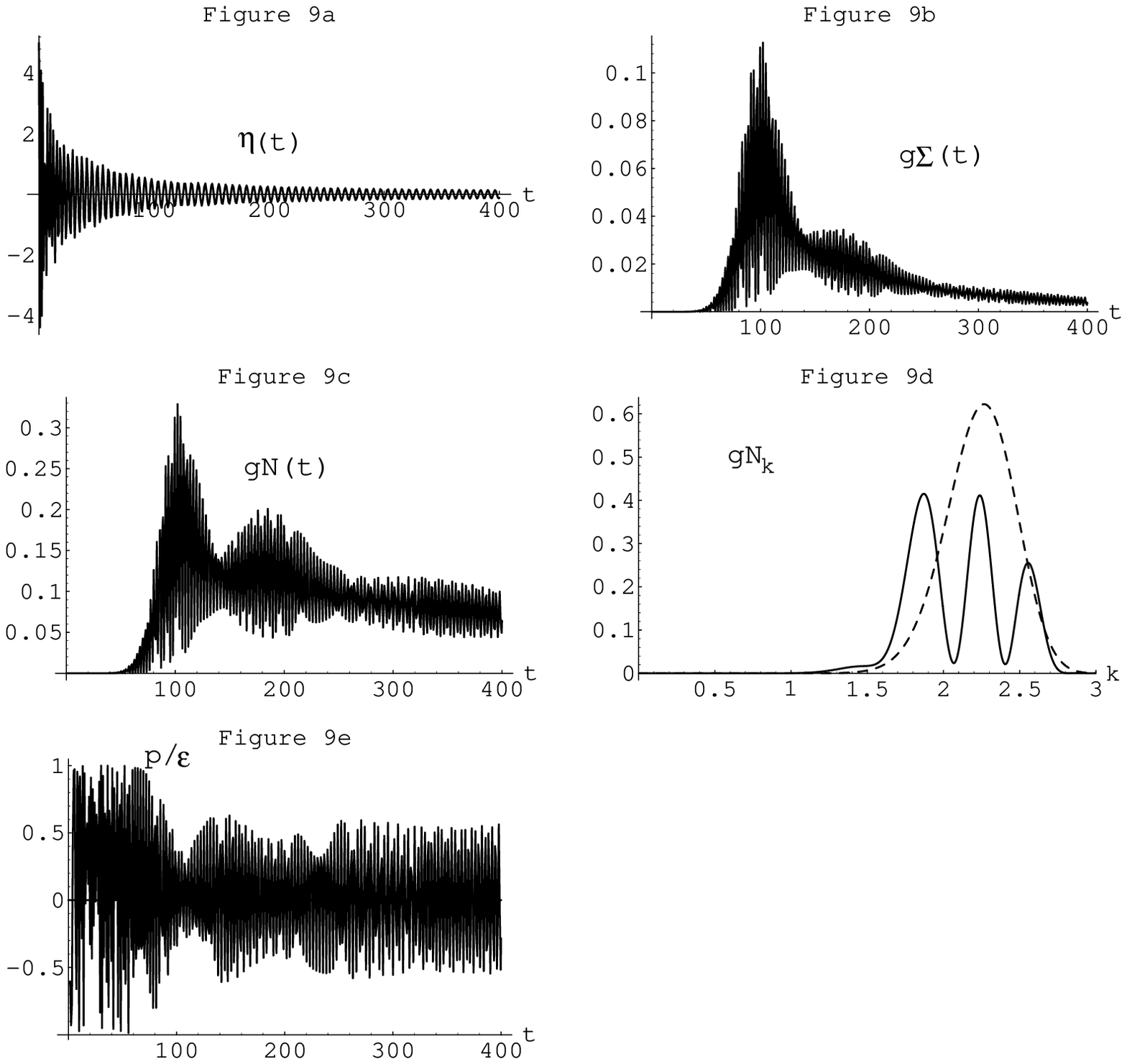}
\caption{Figure 9: Symmetry unbroken, chaotic, large $N$, matter dominated
evolution of (a) the zero mode $\eta(t)$ vs. $t$, (b) the quantum fluctuation
operator $g\Sigma(t)$ vs. $t$, (c) the number of particles $gN(t)$ vs. $t$,
(d) the particle distribution $gN_k(t)$ vs. $k$ at $t=77.4$ (dashed line)
and $t=399.7$ (solid line),  and (e) the ratio of the pressure and energy density
$p(t)/\varepsilon(t)$ vs. $t$ for the parameter values $m^2=+1$, $\eta(t_0) =
5$, $\dot{\eta}(t_0)=0$, $g = 10^{-12}$, $H(t_0) = 0.1$. \label{fig9}}
\end{figure}

\begin{figure}
\epsfig{file=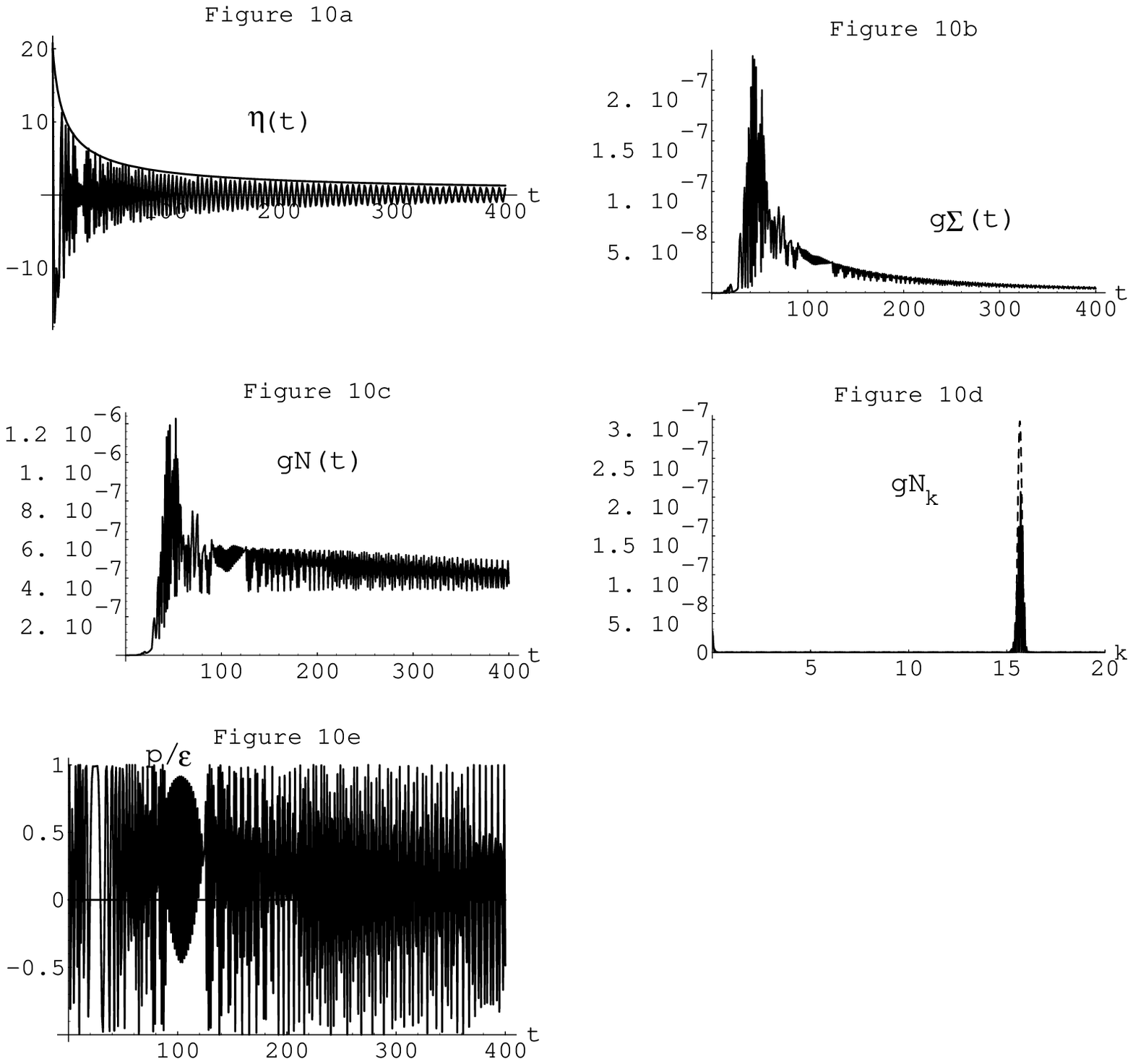}
\caption{Figure 10: Symmetry unbroken, chaotic, Hartree, matter dominated
evolution of (a) the zero mode $\eta(t)$ vs. $t$, (b) the quantum fluctuation
operator $g\Sigma(t)$ vs. $t$, (c) the number of particles $gN(t)$ vs. $t$,
(d) the particle distribution $gN_k(t)$ vs. $k$ at $t=50.5$ (dashed line)
and $t=391.2$ (solid line),  and (e) the ratio of the pressure and energy density
$p(t)/\varepsilon(t)$ vs. $t$ for the parameter values $m^2=+1$, $\eta(t_0) =
12\cdot 3^{1/2}$, $\dot{\eta}(t_0)=0$, $g = 10^{-12}$, $H(t_0) =
0.1$. \label{fig10}}
\end{figure}

\begin{figure}
\epsfig{file=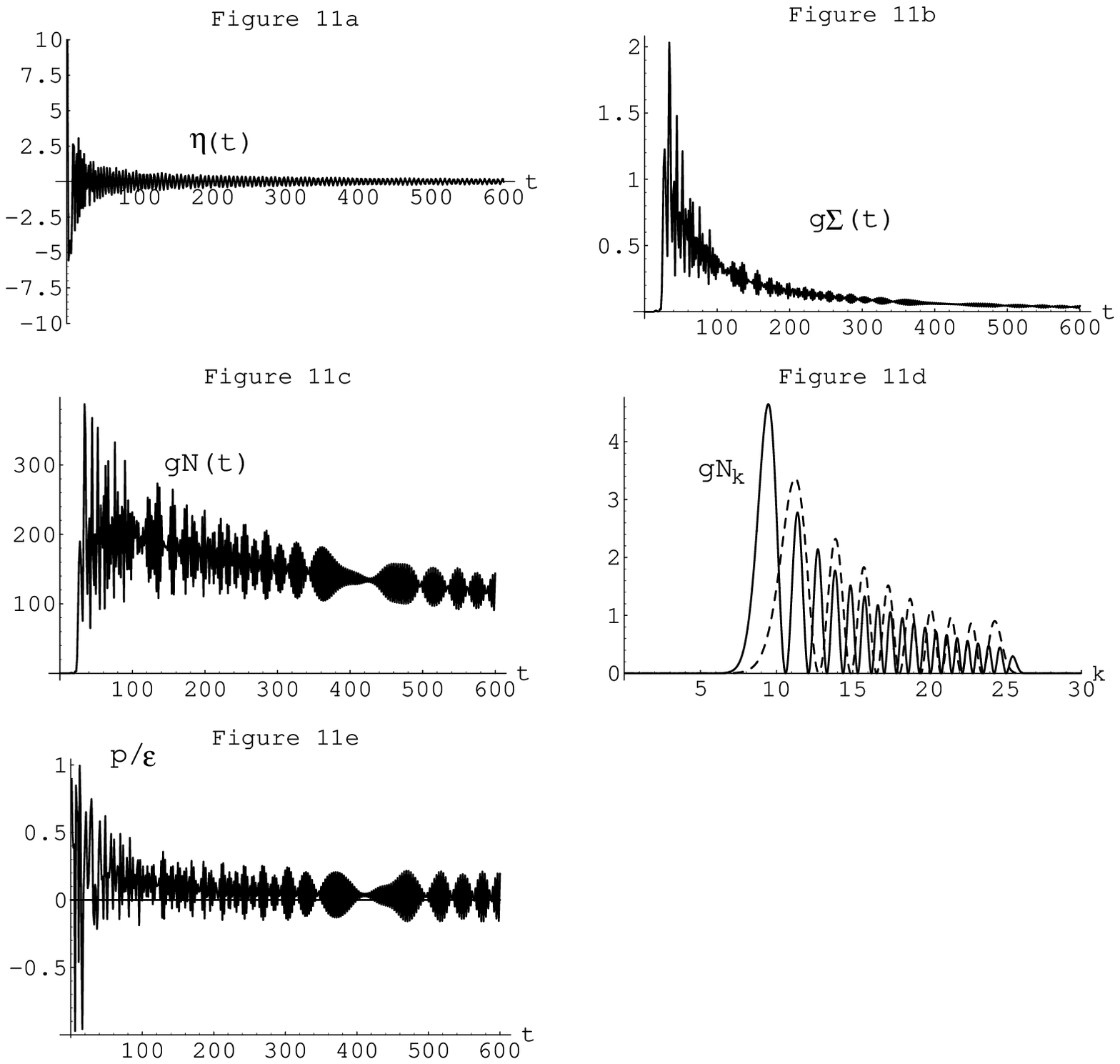}
\caption{Figure 11: Symmetry unbroken, chaotic, large $N$, radiation dominated
evolution of (a) the zero mode $\eta(t)$ vs. $t$, (b) the quantum fluctuation
operator $g\Sigma(t)$ vs. $t$, (c) the number of particles $gN(t)$ vs. $t$,
(d) the particle distribution $gN_k(t)$ vs. $k$ at $t=117.3$ (dashed line)
and $t=393.6$ (solid line),  and (e) the ratio of the pressure and energy density
$p(t)/\varepsilon(t)$ vs. $t$ for the parameter values $m^2=+1$, $\eta(t_0) =
40$, $\dot{\eta}(t_0)=0$, $g = 10^{-12}$, $H(t_0) = 5.0$. \label{fig11}}
\end{figure}

\begin{figure}
\epsfig{file=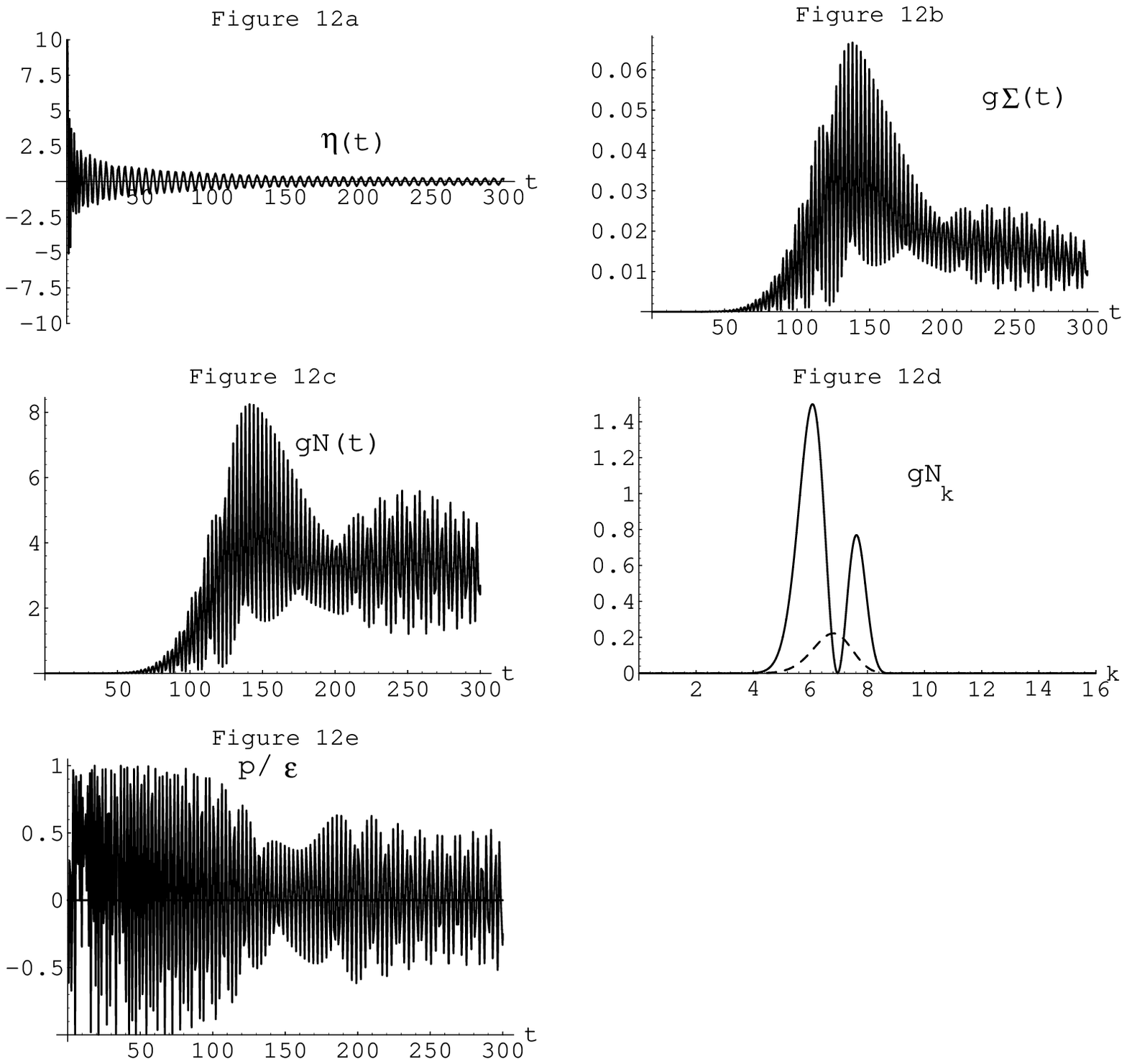}
\caption{Figure 12: Symmetry unbroken, chaotic, large $N$, radiation dominated
evolution of (a) the zero mode $\eta(t)$ vs. $t$, (b) the quantum fluctuation
operator $g\Sigma(t)$ vs. $t$, (c) the number of particles $gN(t)$ vs. $t$,
(d) the particle distribution $gN_k(t)$ vs. $k$ at $t=102.1$ (dashed line)
and $t=251.6$ (solid line),  and (e) the ratio of the pressure and energy density
$p(t)/\varepsilon(t)$ vs. $t$ for the parameter values $m^2=+1$, $\eta(t_0) =
16$, $\dot{\eta}(t_0)=0$, $g = 10^{-12}$, $H(t_0) = 2.0$. \label{fig12}}
\end{figure}

\end{document}